\documentclass{emulateapj}
\usepackage{amssymb}
\usepackage{natbib}
\usepackage{graphicx}
\usepackage{amsmath}

\def\N1Mpc{$N_{\rm 1Mpc}$\/ }
\def\L200{$L_{200}$\/}
\def\n200{$N_{200}^{\rm gal}$\/}
\def\LBCG{$L_{\rm BCG}$\/}

\long\def\symbolfootnote[#1]#2{\begingroup%
\def\thefootnote{\fnsymbol{footnote}}\footnote[#1]{#2}\endgroup} 

\shorttitle{Southern Cosmology Survey I}
\shortauthors{Menanteau et al.}
\submitted{Accepted for publication in ApJ}

\begin{document}

\title{Southern Cosmology Survey I: Optical Cluster Detections and Predictions for the Southern Common-Area Millimeter-Wave Experiments }

\author{Felipe Menanteau\altaffilmark{1},
John P. Hughes\altaffilmark{1},
Raul Jimenez\altaffilmark{2,3}, 
Carlos Hernandez-Monteagudo\altaffilmark{4},
Licia Verde\altaffilmark{2,3},
Arthur Kosowsky\altaffilmark{5},
Kavilan Moodley\altaffilmark{6},
Leopoldo Infante\altaffilmark{7}
and
Nathan Roche\altaffilmark{8}
}

\altaffiltext{1}{Rutgers University, Department of Physics \& Astronomy, 136 Frelinghuysen Rd, Piscataway, NJ 08854, USA }
\altaffiltext{2}{ICREA \& Institute of Space Sciences(CSIC-IEEC), Campus UAB, Bellaterra, Spain}
\altaffiltext{3}{Department of Astrophysical Sciences, Peyton Hall, Princeton University, Princeton, NJ 08544, USA}
\altaffiltext{4}{Max Planck Institut fur Astrophysik, Karl Schwarzschild Str.1,D-85741, Garching Munchen, Germany }
\altaffiltext{5}{University of Pittsburgh, Physics \& Astronomy Department, 100 Allen Hall, 3941 O'Hara Street, Pittsburgh, PA 15260, USA}
\altaffiltext{6}{University of KwaZulu-Natal, Astrophysics \& Cosmology Research Unit, School of Mathematical Sciences, Durban, 4041, South Africa.}
\altaffiltext{7}{Pontificia Universidad Cat\'olica de Chile, Departamento de Astronom\'{i}a, Santiago, Chile}
\altaffiltext{8}{University of Pennsylvania, Physics and Astronomy, 209 South 33rd Street, Philadelphia, PA 19104, USA}

\begin{abstract}

We present first results from the Southern Cosmology Survey, a new
multiwavelength survey of the southern sky coordinated with the
Atacama Cosmology Telescope (ACT), a recently commissioned
ground-based mm-band Cosmic Microwave Background experiment.  This
article presents a full analysis of archival optical multi-band
imaging data covering an 8 square degree region near right ascension
23 hours and declination -55 degrees, obtained by the Blanco 4-m
telescope and Mosaic-II camera in late 2005. We describe the pipeline
we have developed to process this large data volume, obtain accurate
photometric redshifts, and detect optical clusters. Our cluster
finding process uses the combination of a matched spatial filter,
photometric redshift probability distributions and richness
estimation. We present photometric redshifts, richness estimates,
luminosities, and masses for 8 new optically-selected clusters with
mass greater than $3\times10^{14}M_{\sun}$ at redshifts out to 0.7. We
also present estimates for the expected Sunyaev-Zel'dovich effect
(SZE) signal from these clusters as specific predictions for upcoming
observations by ACT, the South Pole Telescope and Atacama Pathfinder
Experiment.

\end{abstract}


\section{Introduction}

The new generation of high-angular resolution Cosmic Microwave
Background (CMB) ground-based experiments represented by the the
Atacama Cosmology Telescope (ACT) \citep{Kosowsky06,Fowler07} and the
South Pole Telescope (SPT) \citep{SPTref} are currently targeting
their observations in a common area in the southern sky that will
ultimately cover several hundreds to thousands of square degrees.  These experiments
will provide a blind survey of the oldest light in the Universe at
wavelengths of $1-2$~mm and angular scales beyond the resolution
limits of the WMAP and Planck satellites. At these arcminute angular scales,
temperature fluctuations in the CMB are
dominated by secondary effects arising from the formation of
large-scale structure in the universe. One of the strongest effects is
the imprint left by galaxy clusters though the Sunyaev Zel'dovich
effect (SZE) \citep{SZ80} in which CMB photons suffer inverse Compton
scattering by the hot intracluster gas.  ACT and SPT are designed to
detect the SZE, through its frequency-dependence: these experiments
will measure temperature shifts of the CMB radiation corresponding to
a decrement below and an increment above the ``null'' frequency around
220~GHz.

Much can be learned about the Universe from these surveys.  
First,  accurate systematics-free maps will allow 
measurement of the primary
power spectrum of temperature fluctuations at all scales on which they
are the dominant contribution. 
Second, these data sets will result in a complete census of massive clusters
to arbitrarily large distances, limited only by a minimum cluster mass
set largely by the instrumental sensitivity and expected to be several
$10^{14}\, M_\odot$ \citep{SPTref,Sehgal07}. 
Thanks to the relatively clean
selection function as well as the redshift independence of the SZE,
the cluster sample, especially the evolution of the number density of
clusters with redshift, will be quite sensitive to the growth of
structure in the Universe offering a potentially powerful probe of
dark energy \citep{Carlstrom-02}. Moreover, the SZE data in
combination with optical, UV and X-ray observations can teach us a
great deal about the detailed physics of cluster atmospheres and
galaxy evolution in these dense environments.

Significant observing time and effort has been devoted to the
development of techniques and the detection of galaxy clusters using
large-area optical catalogs and X-ray observations. Several projects
have taken advantage of large-area CCD imaging and have developed
automated cluster detection schemes to produce large catalogs of
clusters of galaxies \citep[see][for
  example]{MaxBCG,Postman96,Postman01,NoSOCSI,NoSOCSII,NoSOCSIII,RCS}
which target the relative over-abundance of galaxies over a range of
redshifts. Similarly, X-ray surveys such as the {\it ROSAT} All Sky Survey
\citep{Ebeling98,Bohringer01,Mullis03} produced catalogs with hundreds
of galaxy clusters, while pointed X-ray observations have discovered
systems up to $z\simeq1.4$ \citep{Mullis05}.

With this article we inaugurate the Southern Cosmology Survey (SCS).
This project, funded by the National Science Foundation under the
Partnership in Research and Education (PIRE) program, is a
multiwavelength (radio, mm-band, optical, UV, and X-ray) large area
survey specifically coordinated with ACT observations of the southern
sky. The goal of the SCS is to maximize the scientific return from
the new ground-based CMB experiments and therefore focuses on specific
observational studies relevant to this science, such as the selection
function of galaxy clusters across wavebands, cluster mass
determination, and the establishment a "gold" sample of clusters for
cosmology and galaxy evolution studies.
Here we present results from an $\simeq8$~deg$^2$ optical imaging
survey of the southern sky that overlaps the common SZ survey
region. The purpose of this paper is twofold: (1) to present the
details of our data reduction pipeline and analysis software and (2)
identify new galaxy clusters, constrain their redshifts and masses,
and predict their SZ signals. Photometric redshifts come from the
4-band imaging data, while our mass estimates are inferred from the
optical luminosity ($L_{200}$) and richness (\n200) of the clusters,
using relations calibrated by the Sloan Digital Sky Survey (SDSS). For
the eight massive clusters, out of 38 identified (37 are new sources)
in the survey area, we present positions, richness estimates, masses,
and predictions for the integrated Compton $y$-distortion of the SZE
using empirical power-law relations based on $N$-body
simulations. Throughout this paper we assume a flat cosmology with
$H_0=100 h$~km~s$^{-1}$~Mpc$^{-1}$, $h=0.7$ and matter density
$\Omega_m=0.3$.

\section{Dataset and Methodology}

Our study is based on the optical multi-band analysis of public data
from the Blanco Cosmology
Survey\footnote{http://cosmology.uiuc.edu/BCS/} (BCS).
This is a NOAO Large Survey Project that was awarded 45 nights over
three years on the Blanco 4-m telescope at the Cerro Tololo
InterAmerican Observatory (CTIO).  The survey aims to image two 50
square-degree patches of the southern sky in four optical bands ({\it
  griz}) using the $8192\times8192$~pixel (0.36~deg$^2$) MOSAIC~II
camera in order to attain a sensitivity about an order of magnitude
deeper than the SDSS imaging.  The targeted areas are centered near
declinations of $-$55$^{\circ}$ and $-$52$^{\circ}$ at right
ascensions of 23 hr and 5 hr respectively; each of these patches lies
within a larger common region of the southern sky that both ACT and
SPT plan to survey. The BCS began in 2005 and has completed
three years of data taking. For this paper we have processed and
analyzed public data from the first year of the survey using
an independent software pipeline developed by us at Rutgers
University. The data we present was obtained on 15 nights of observing
near the end of November and the beginning of December 2005 and cover
an area of $\simeq8$~deg$^2$ in the 23 hr region. In
Table~\ref{tab:obs} we show the observing dates, photometric
conditions, lunar illumination and observed bands for the 19 tiles
that make up the full extent of the observations analyzed here.  In
the following we describe the steps followed and tasks performed by
the pipeline.

\begin{deluxetable}{ccccccc}
\tablecaption{2005 Observations in the 23hr Field}
\tablehead{
\multicolumn{3}{c}{} & \multicolumn{4}{c}{\# of Tiles Obs}\\
\colhead{Date} & 
\colhead{Photometric} & 
\colhead{Lunar Illum} & 
\colhead{$g$} & 
\colhead{$r$} &
\colhead{$i$} &
\colhead{$z$} 
}
\startdata
18 Nov 2005 &   yes &    89.4\% &  1.0 &  1.0 &  3.0 &  3.0 \\
19 Nov 2005 &   yes &     0.0\% &  4.5 &  4.5 &  1.0 &  1.0 \\
20 Nov 2005 &    no &     0.0\% &  8.0 &  8.0 &  0.0 &  0.0 \\
22 Nov 2005 &    no &     0.0\% &  0.0 &  0.0 &  0.0 &  0.0 \\
24 Nov 2005 &   yes &     0.0\% &  2.0 &  2.0 &  0.0 &  0.0 \\
26 Nov 2005 &   yes &     0.0\% &  3.0 &  3.0 &  1.0 &  1.0 \\
28 Nov 2005 &   yes &     0.0\% &  0.0 &  0.0 &  2.7 &  2.7 \\
30 Nov 2005 &   yes &     0.0\% &  0.0 &  0.0 &  0.3 &  0.3 \\
02 Dec 2005 &   yes &     0.0\% &  0.5 &  0.5 &  0.0 &  0.0 \\
04 Dec 2005 &    no &    15.3\% &  0.0 &  0.0 &  0.0 &  0.0 \\
05 Dec 2005 &    no &    24.5\% &  0.0 &  0.0 &  1.3 &  1.7 \\
06 Dec 2005 &   yes &    35.1\% &  0.0 &  0.0 &  1.3 &  1.0 \\
08 Dec 2005 &   yes &    57.8\% &  0.0 &  0.0 &  2.3 &  2.7 \\
10 Dec 2005 &   yes &    78.4\% &  0.0 &  0.0 &  3.0 &  3.0 \\
11 Dec 2005 &   yes &    86.6\% &  0.0 &  0.0 &  3.0 &  3.3 \\
\enddata
\label{tab:obs}
\tablecomments{Observing conditions during the 2005 run of the BCS,
  consisting of only the 19 tiles that were fully observed in all 4
  bands in the 23hr region. Lunar illumination is the percentage at
  midnight local time in the direction toward the center of the region
  surveyed (R.A. 23 hours, decl. $-$55.2deg).}
\end{deluxetable}

\subsection{The Rutgers Southern Cosmology Pipeline}

The Rutgers Southern Cosmology image analysis pipeline is written in
Python with a scalable object-oriented design based on existing public
astronomical software that is aimed at processing a large dataset in a
repeatable, stable and semi-automated fashion.

The initial standard image processing steps for each observing night
are handled by the IRAF/mscred \citep{Valdes98} procedures via the
STScI/Pyraf interface. These include: overscan trim, bias correction,
CCD cross-talk coefficients corrections as well as dome flat field
correction. The pipeline also executes secondary CCD calibration steps
on the science images which include the creation of super sky-flats,
fringe patterns for $i-$ and $z-$bands and their corresponding
correction and removal. Additionally, procedures affecting the
cosmetic appearance of the images, such as cosmic ray rejection,
removal of saturated star bleed-trails, and generation of bad pixel
masks, are automatically performed at this stage. Astrometric
re-calibration and WCS plate solution are also handled automatically
at this stage on the pre-stack science images using IRAF's
mscred/mscmatch task by matching several hundred sources within each
tile with stars from the US Naval Observatory Catalog. We achieve good
accurate astrometric solutions (the residual error in matched source
positions was typically $<0.1''$) as tested using the overlapping
regions between neighboring pointings. Photometric standard star
fields were processed together with the normal science images and
photometric zero points for each observing night were obtained using a
few hundred standards from the Southern Hemisphere Standards Stars
Catalog \citep{smith07}. Like these authors we use the AB magnitude
system.

The survey strategy followed a predetermined observing pattern, which
typically consisted of exposures of $2\times125s$, $2\times300s$,
$3\times450s$ and $3\times235s$ in the $g,r,i$ and $z-$bands
respectively with offsets of $3-5$~arcmin (within each filter) intended
to provide significant overlap between neighboring MOSAIC~II tiles and
fill in the gaps between CCDs chips. We used the overlapping regions between
tiles to adjust the photometric zero points of non-photometric nights using 
matched sources from
adjacent photometric tiles.  This ensured a homogeneous photometric
calibration across the full survey region with typical variations 
below $0.02$~mags.

Image alignment, stacking and combination as well as catalog
generation are performed at a secondary stage by the pipeline using
association files, which describes a logical group of exposures and
filters, created for each tile.  Science images were mosaiced, aligned
and median combined using SWarp \citep{SWarp} to a plate scale of
$0.266''$/pixel.  Source detection and photometry measurements for the
science catalogs were performed using SExtractor \citep{SEx} in
dual-image mode in which sources were identified on the $i-$~band
images using a $1.5\sigma$ detection threshold, while magnitudes were
extracted at matching locations from all 4 bands.

\begin{figure}
\includegraphics[width=\columnwidth]{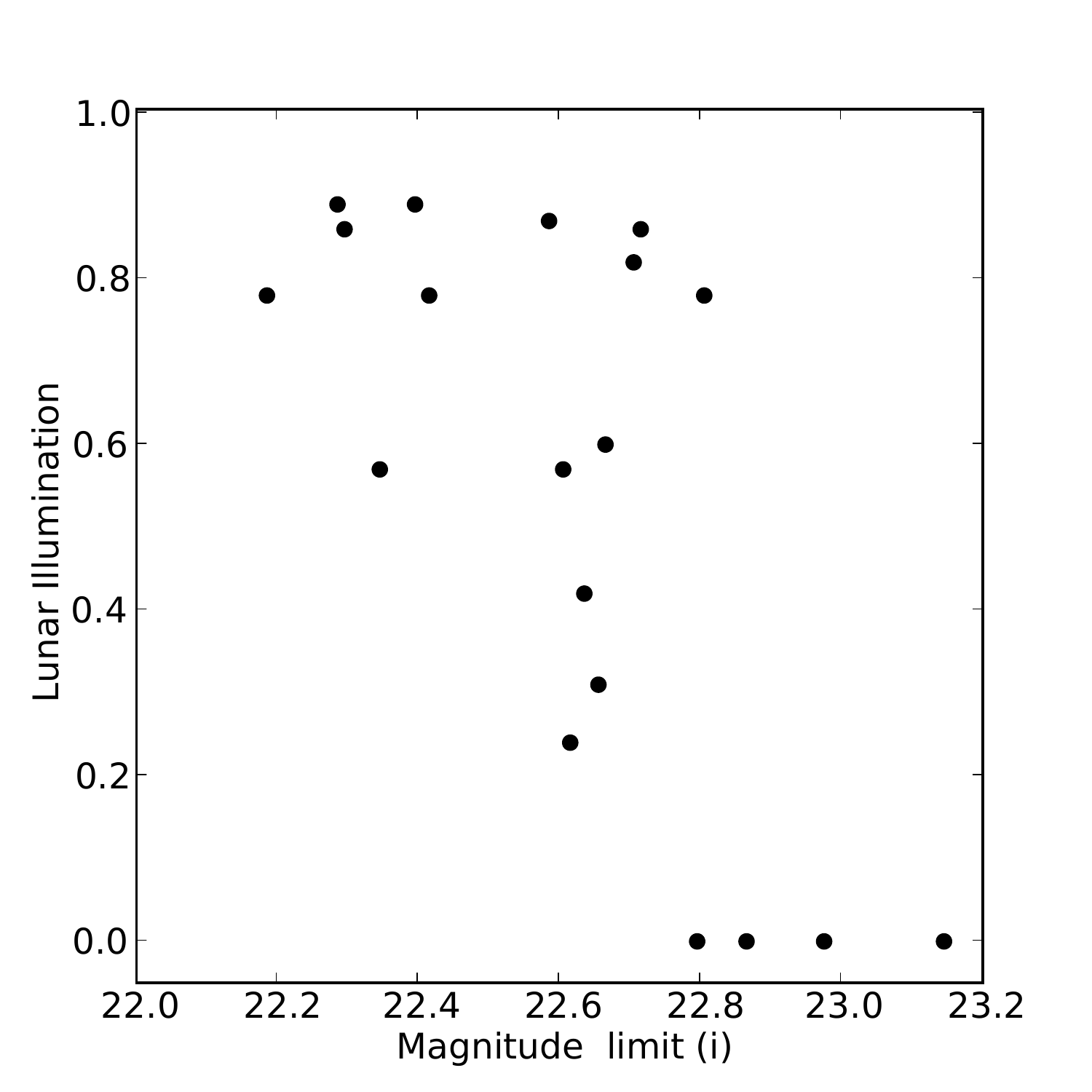}
\caption{The mean lunar illumination fraction for each tile at time of 
  observation in the
  23hr field as a function of the $i$-band magnitude limit for 90\%
  completeness.}
\label{fig:illum}
\end{figure}
 
As our data set is composed of observations taken over several weeks
under varying conditions during the 2005 campaign, we determined, for
each tile individually, the $i$-band magnitude limit at which the
galaxy detection limit was complete to $90\%$. To compute this limit
we use the fact that the galaxy number counts follow a power-law
function, which we fitted in the magnitude range $19.5<i<21.5$ in each
tile and extrapolated to obtain the magnitude at which the galaxy
number counts dropped by $10\%$.  We took this as the $90\%$
completeness limit for the tile. We found variations of roughly 1
magnitude on the limits among the 19 tiles and, in an attempt to
understand this, we investigated a possible correlation with lunar
illumination at the time of observation.  Figure~\ref{fig:illum} shows
a clear trend between the $i$-band magnitude limit and the lunar
illumination.  We report a mean limit $i = 22.62 \pm 0.25$ and we set
a conservative magnitude limit of $i=22.5$ for our full catalog.

\subsection{Photometric Redshifts}

\begin{figure}[h!]
\includegraphics[width=3.1in]{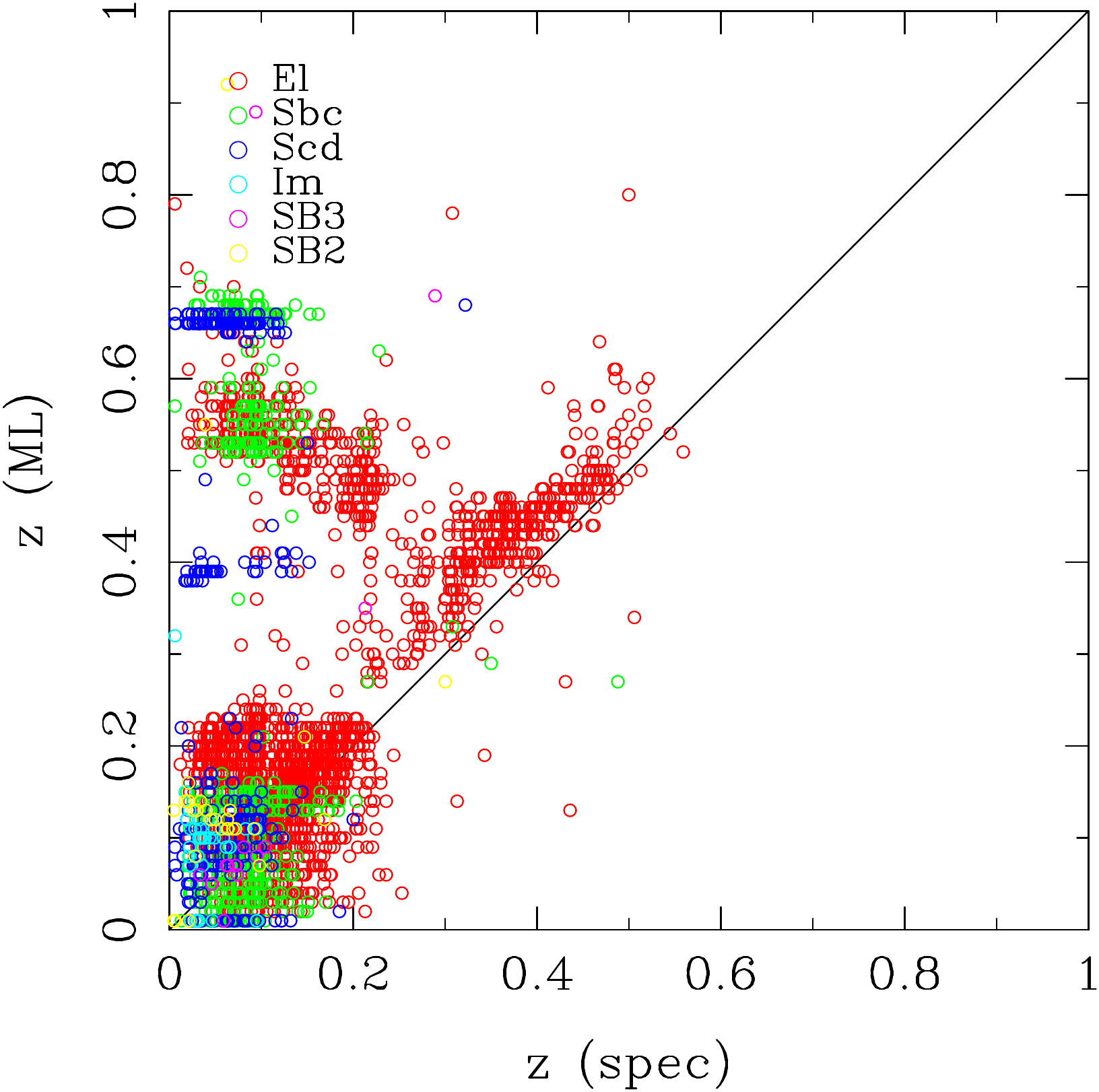}\\
\includegraphics[width=3.1in]{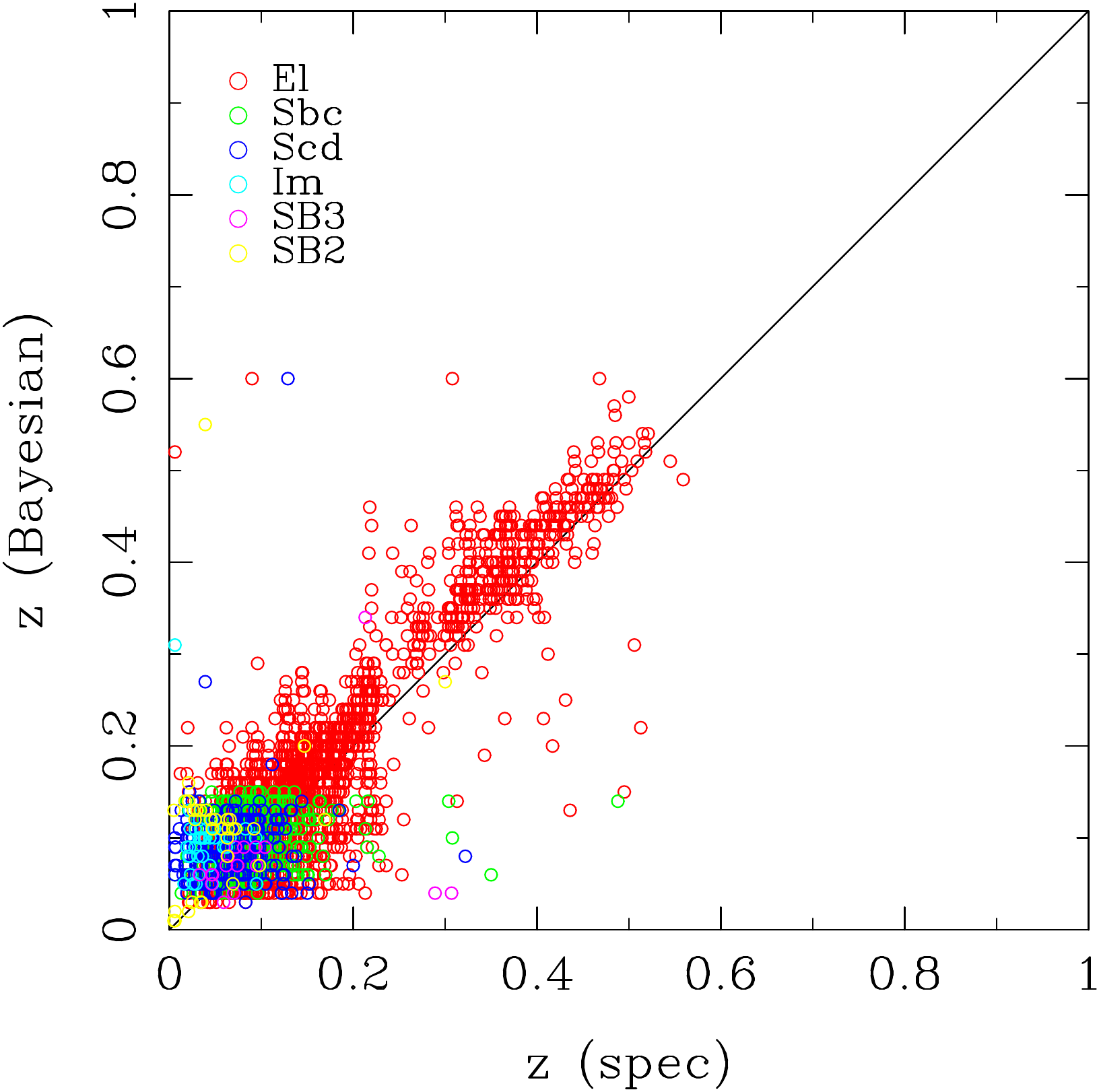}
\caption{Comparison between photometric and spectroscopic redshifts
  for 5000 galaxies in the SDSS DR6. The upper panel shows the
  comparison between spectroscopic and photometric redshifts obtained
  under the maximum likelihood assumption. The lower panel uses a
  Bayesian assumption with a custom empirical prior on galaxy
  brightness for the photometric redshifts. Symbols are color coded 
  according to
  the best determined SED by BPZ. SEDs for E/S0 galaxies tend to
  dominate at higher redshifts as the SDSS is biased towards
  early-type galaxies which are the most luminous population at these
  redshifts.}
\label{fig:bpz}
\end{figure}

>From the multi-band photometry the pipeline computes photometric
redshifts and redshift probability distributions $p_{\rm BPZ}(z)$ for
each object using the $g,r,i,z$ isophotal magnitudes, as defined by
the i-band detection, and the BPZ code \citep{BPZ}.  We use a
magnitude-based empirical prior (Benitez, private communication) taken
from the SDSS and HDF-N spectroscopic redshift distributions, which
accounts for the tendency of fainter galaxies to be more likely found
at higher redshifts \citep[See Fig~4. from][as an
  example]{BPZ}. Because the area covered by the available NOAO
imaging does not include any publicly available spectroscopic redshift
information for $z>0.1$ (NASA/IPAC Extragalactic Database, NED), we
investigated the accuracy of our photometric redshift estimates using
ancillary data. As the ability of BPZ to estimate photometric
redshifts at fainter magnitudes ($i>20$) from multi-band photometry
has been consistently established in the past using filter sets
similar to ours here \citep[for example]{BPZ04, Mobasher04, Cross04},
we focused on the redshift accuracy at $z<0.5$.  To this end we
extracted $g,r,i,z$ photometry from the DR6 SDSS for 5000
randomly-selected bright galaxies ($r<20$~mag) with reliable
spectroscopic redshifts, matching the depth and signal-to-noise ratio
of our galaxy sample.  We computed photometric redshifts for the SDSS
spectroscopic sample in the same way as just described and compared
the resulting values to the spectroscopic redshifts.  We found, not
surprisingly, that simply employing the maximum-likelihood (ML)
condition is an ill-suited approach for redshifts below $z<0.3$ in the
absence of a bandpass bluer than 3000\AA\/ as it largely
over-estimates redshifts and produces an unacceptable number of
catastrophic outliers. Recently, \cite{Niemack08} have demonstrated
how the addition of bluer bands using GALEX UV imaging greatly
improves ML estimates and reduces the need for priors.
On the other hand, Bayesian estimates give results with typical rms
errors of $\delta z\sim0.02$ and with almost no catastrophic
outliers. In Fig.~\ref{fig:bpz} we show the results of our comparison
between ML and Bayesian photometric versus spectroscopic redshifts
color-coded according to the spectral energy distribution (SED)
determined by the BPZ code.  It is clear from the figure that at
higher redshifts the SDSS population is dominated by early types as
these tend to be the most luminous objects.
We also note (see fig.~\ref{fig:bpz}, lower panel) that on average our
photometric redshifts tend to overpredict the true redshifts. The mean bias
level $\langle z_{\rm spec}-z_{\rm BPZ} \rangle$ for galaxies with
E/S0s SEDs is largest around $z_{\rm spec}\sim 0.3$ where it is on the
order of $\delta z\sim -0.03$. In table~\ref{tab:bpz} we show the mean
bias and standard deviation for three redshift intervals for $z_{\rm
  spec}-z_{\rm BPZ}$ as well as the standard $dz$ defined as $dz =
z_{\rm spec}-z_{\rm BPZ}/(1+z_{\rm spec})$.
In summary we are able to determine the redshifts for early-type
galaxies to an accuracy better than 0.1 across the redshift range of
the survey.  This is encouraging since early-type galaxies are the
predominant population in clusters of galaxies and good photometric redshift
determination is essential for successful cluster finding, as we
discuss in the next section.

\begin{figure}
\includegraphics[width=\columnwidth]{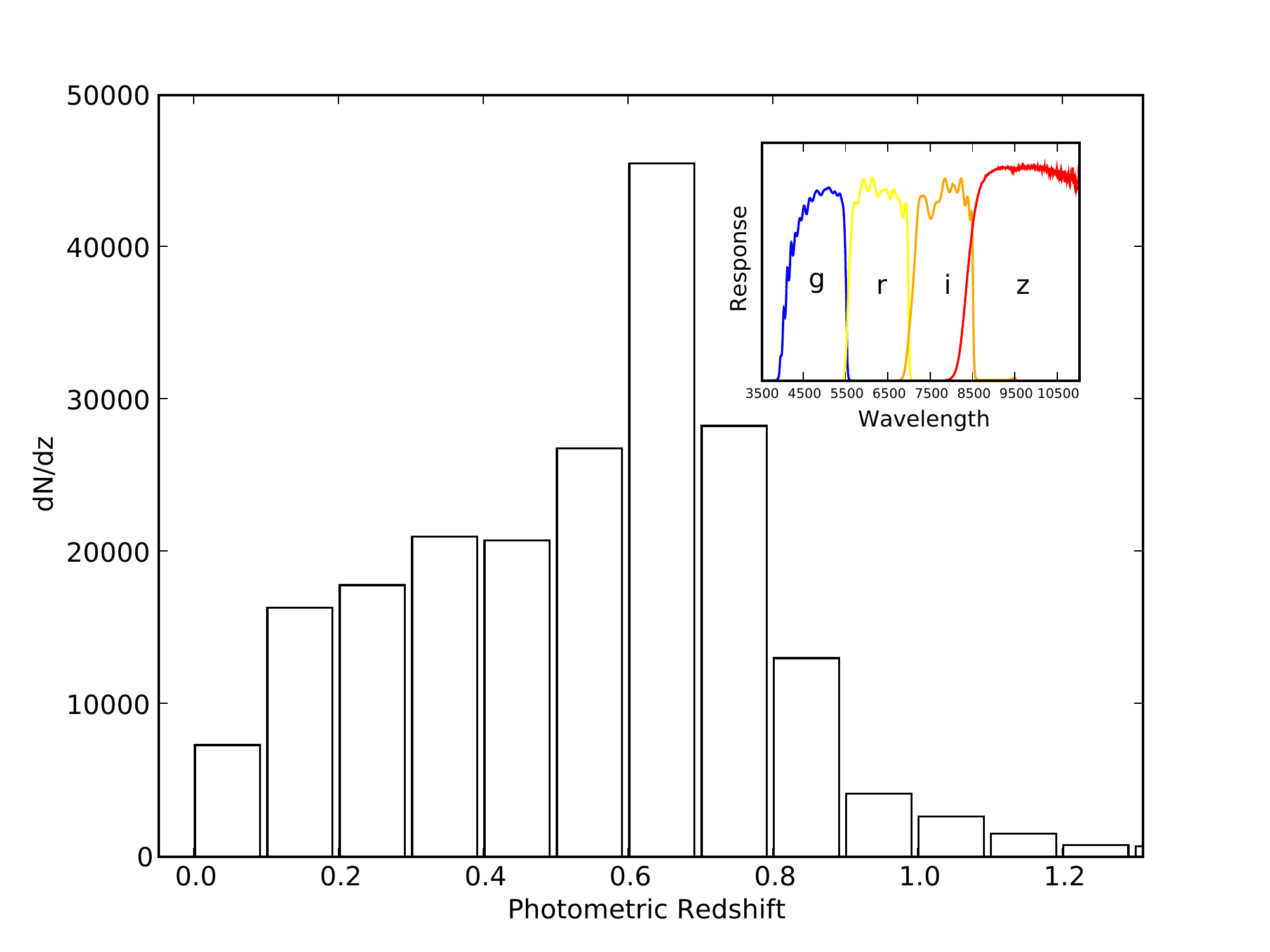}
\caption{The photometric redshift distribution of the galaxy sample
  used for finding clusters (i.e.: galaxies $i<22.5$~mag).}
\label{fig:dndz}
\end{figure}

In Figure~\ref{fig:dndz} we show the photometric redshift distribution for
all galaxies within our flux completeness limit, $i<22.5$, as well as
the filter responses for the
survey\footnote{http://www.ctio.noao.edu/instruments/FILTERS/index.html}.
Our distribution peaks around $z\sim0.6$, which sets a conservative
upper limit to the redshift at which we are able to detect optical
clusters.

\begin{deluxetable}{ccccc}
\tablecaption{Photometric Redshift Simulations Statistics}
\tablehead{
\colhead{Redshift} & 
\colhead{$\langle z_{\rm spec}-z_{\rm BPZ} \rangle$} & 
\colhead{$(z_{\rm spec}-z_{\rm BPZ})_{\rm rms}$} &
\colhead{$\langle dz \rangle$} &
\colhead{$\sigma_z$} }
\startdata
$0.0-0.2$ & -0.017 &  0.042 & -0.015 &  0.038\\
$0.2-0.4$ & -0.027 &  0.059 & -0.020 &  0.047\\
$0.4-0.6$ & -0.002 &  0.070 & -0.001 &  0.048\\
\enddata
\label{tab:bpz}
\tablecomments{The mean difference and standard deviation between
  spectroscopic and recovered photometric redshifts as well as for $dz$ for all galaxies with
  SED determined to be E/S0s in three redshift ranges.}
\end{deluxetable}

\subsection{Computing Overdensities and Finding Clusters}

One of the main goals of the current SZE experiments is to define a
mass-selected sample of galaxy clusters out to large redshifts. At
long last this is beginning to happen \citep{SPT08, Menanteau-Hughes-09},
after a number of successful individual detections of the SZE in well-known
optical or x-ray clusters \citep[see][and references
therein]{Birkinshaw91,SZE4,SZE5,SZE2,SZE3,SZE1,SZE6}.
If we want to
understand the systematics of SZE surveys it is crucial to compare
with cluster identifications using independent methods. In this
section, we describe our effort to select clusters of galaxies from
multi-wavelength optical imaging. There are several methodologies and
a plethora of papers describing these techniques \citep[for
  example]{Postman96,RCS,MaxBCG,eisen08} but they all rely on the same
well-known properties of galaxy clusters: a) early-type galaxies are
the dominant population, b) cluster galaxies have very similar colors,
and display tight color-magnitude relationships across several orders
of magnitude in luminosity, and c) the surface number density of
cluster galaxies falls off with distance from the center roughly as a
power law $P(r)\propto1/r^\alpha$. We search for clusters using a
matched filter approach similar to the one described in
\cite{Postman96} and then define membership and estimate richness of
the clusters using the MaxBCG prescription \citep{MaxBCG}.

\begin{figure}
\centerline{\includegraphics[width=\columnwidth,angle=0]{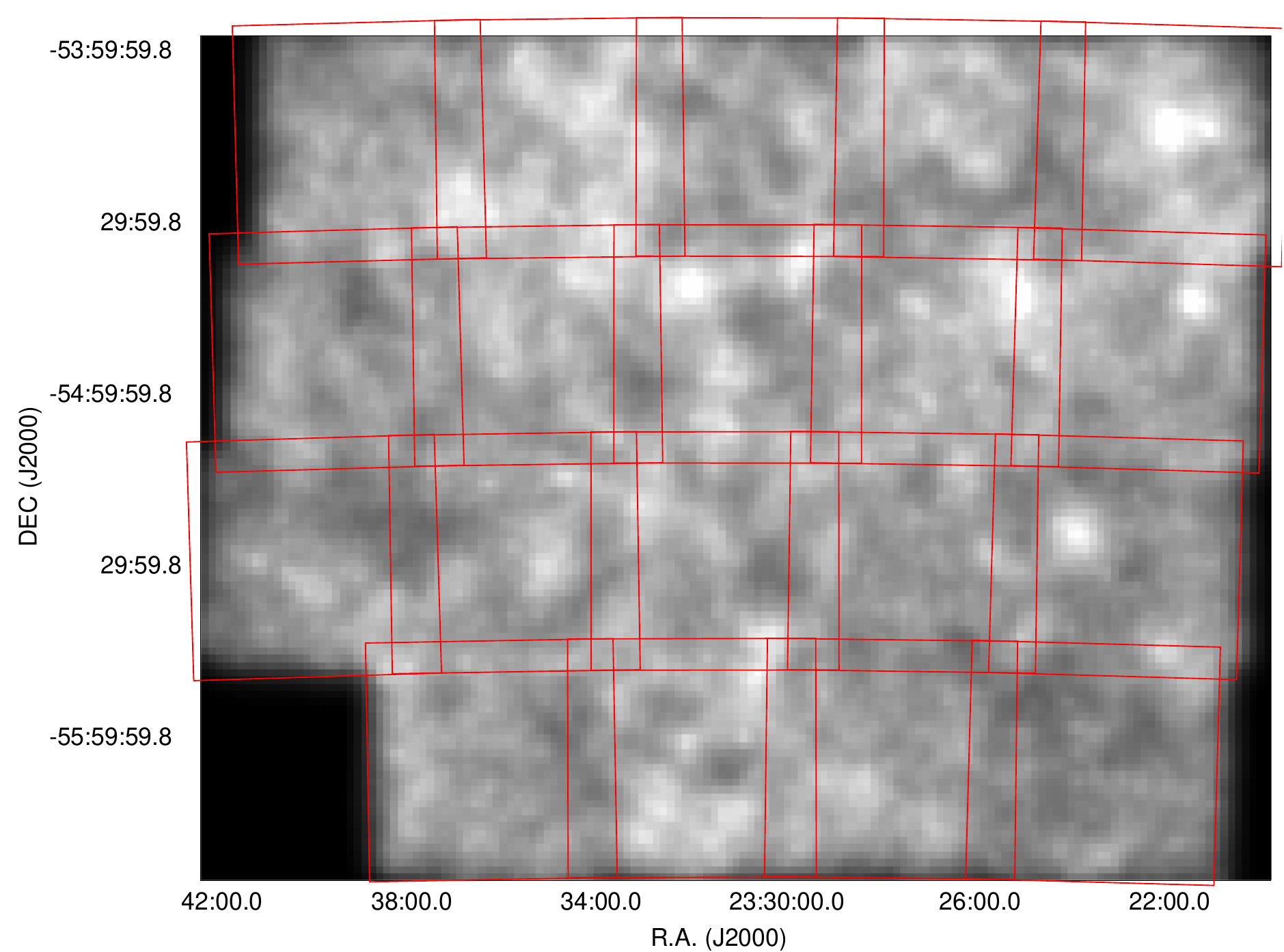}}
\caption{The computed likelihood density map image centered at $z=0.2$
  and width $\Delta z=0.1$ over the 23hr field. Bright regions in the
  image represent denser areas. The red lines represent the area
  covered by each of the 19 tiles that comprise the area studied.}
\label{fig:dense}
\end{figure}

Our cluster finder method folds in the contributions from a cluster
spatial profile filter function $P(r)$, a luminosity weight $L(m)$ and
the BPZ redshift probability distribution $p_{\rm BPZ}(z)$ from each
source to generate likelihood density maps (at pixel positions
denoted by $i,j$) or a ``filtered'' galaxy catalog $S(i,j)(z)$ over
the area covered by the survey as a function of redshift, namely,
\begin{equation}
S(i,j)(z) = \sum_{k=1}^{\rm N_g} P(r_k[i,j])~L(m_k) \int_{z-\Delta z}^{z+\Delta z} p_{\rm BPZ}(z_k)dz.
\end{equation}
Specifically we use a profile with the form
\begin{eqnarray}
P(r/r_c) &=& \frac{1}{\sqrt{1+(r/r_c)^2}} - \frac{1}{\sqrt{1+(r_{cut}/r_c)^2}}, {\rm ~if~} r<r_{\rm cut}  \nonumber\\
         &=& 0, {\rm otherwise} \nonumber \\
\end{eqnarray}
which is normalized as
\begin{equation}
\int_0^{\infty} P(r/r_c)2\pi r dr = 1
\end{equation}   
and where $r_c$ is the typical cluster core radius and $r_{\rm cut}$
is the cutoff limit for the function. In our analysis we chose $r_c=175$~kpc and
$r_{\rm cut} = 10r_c$. We also use a luminosity weight $L(m)$ given by
\begin{equation}
L(m) = \frac{\phi(m-m^*)10^{-0.4(m-m^*)}}{b(m)} = \frac{\Phi(m-m^*)}{b(m)}
\end{equation}
where $m^*$ is the apparent magnitude corresponding to $M^*$. This
function is normalized as
\begin{equation}
 \int_0^{m_{\rm lim}} \Phi(m-m^*)dm = 1
\end{equation}
where $m_{\rm lim}$ is the flux limit of the sample ($i=22.5$), $b(m)$
is the number of background galaxies and $\phi(m)$ is the
\citet{Schechter-76} galaxy luminosity function. We use the parameters
computed from \cite{Brown-07} for the evolving luminosity function of
red galaxies, with a faint-end slope $\alpha=-0.5$ and $M^*(z)$
between $0<z<1$. For our estimation of $b(m)$ we use the number counts
from \cite{Yasuda-01}. We generate likelihood density maps with a
constant pixel scale of $1.2$~arcmin at $\Delta z=0.1$ intervals
between $0.1<z<0.8$ over the surveyed regions. In
Figure~\ref{fig:dense} we show an example of a likelihood density map
centered at $z=0.2$ on which we superpose outlines of the  19
tiles that define the region studied here.

Cluster candidates are selected from the peaks of the likelihood
density maps. In order to define peaks consistently we constructed
noise maps by randomizing the positions of the input catalog to
produce likelihood density maps following exactly the same procedure
as described above. These maps represent the noise floor level above
which we desire to detect clusters. We define our initial list of
cluster candidates from $2\sigma$ peaks in the likelihood density
maps, where $\sigma$ is defined as the median value in the noise
maps. 
Cluster candidates were checked for multiple detection in
different likelihood maps. To avoid duplication we
considered a system unique if detected in two adjacent redshift maps 
and with the same center (i.e., within 3$^\prime$).

\subsection{Contamination and Completeness}

We perfomed simulations to investigate our cluster selection function
by estimating the contamination and recovery rates of our cluster
finding technique. The lack of distance information in imaging surveys
is the principal source of contamination as fluctuations in the
projected two-dimensional galaxy distribution as well as random
alignments of poor groups may result in false apparent
overdensities. We explore this issue following the same methodology as
used by \cite{Postman02} and \cite{NoSOCSII} which rely on
generating Monte Carlo representations of the galaxy sample
with an angular two-point correlation function similar to that
observed \citep{Infante94}. As described in \cite{Postman02} we
implemented the Rayleigh-L\'evy (RL) random walk process using 
Mandelbrot's (1975) elegant fractal prescription  to simulate galaxy
positions on the sky, such that galaxy pairs are placed in a randomly
chosen direction at distance $\theta$ drawn from the distribution:
\begin{equation}
P(>\theta) =
 \begin{cases}
 (\theta / \theta_{0})^{-d}, & \mbox{if }\theta\ge\theta_0 \\
  1,                       & \mbox{if }\theta<\theta_0,  \\
\end{cases}
\end{equation}
where we chose $\theta_0$ and $d$ to match the observed galaxy
distribution of our sample. In practice we generate simulated
distributions by starting from a randomly selected location within the
survey boundaries and generate positions following the RL random walk
allowing up to seven galaxies to be drawn around this location. We
then select a new center randomly and the process is repeated until we
generate the same number of galaxies as in the observed sample. We
then process the RL distribution to generate likelihood maps using the
same procedure and parameters as for the real data and use these to
investigate the rate of false detections as a function of estimated
redshift. 
Since the RL distributions by construction do not explicitly include
clusters, we assess the false positive cluster detection fraction by
taking the ratio of detections in the simulations per area unit to the
observed number of candidates in the real data. We find that at low
redshift the false positive fraction is zero (there is virtually no
contamination), while at redshifts of $z=0.6$ and $z=0.7$ the false
positive fraction grows to values of 1\% and 19\%, respectively. We
conclude that false positives are not an important source of spurious
detections.

\begin{figure}
\centerline{
\includegraphics[width=3in]{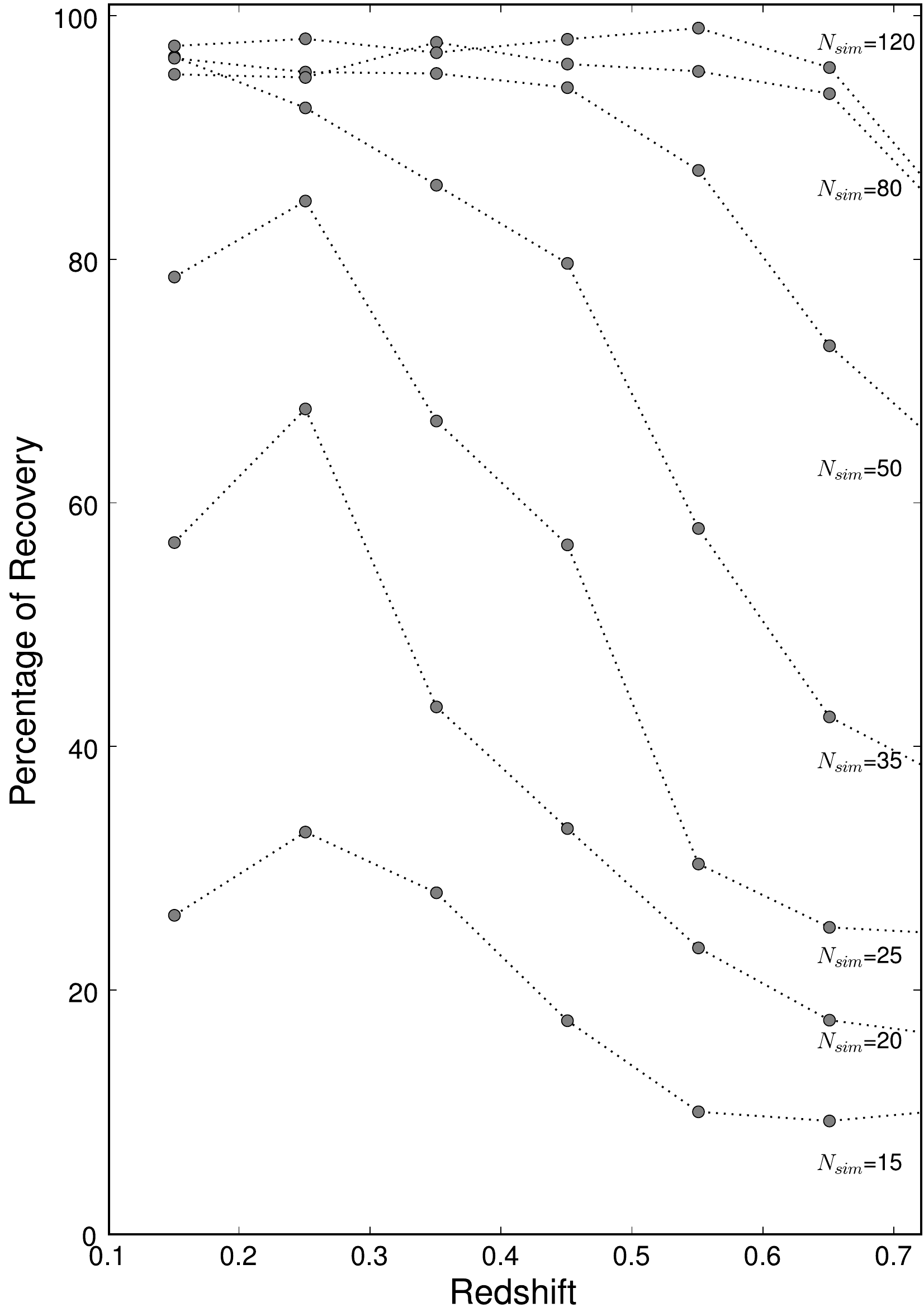}}
\caption{The cluster recovery fraction as a function of redshift as
  extracted from the simulations for clusters with $N_{sim}=15, 20,
  25,35,50,80$ and $120$ galaxies and a profile with slope
  $\alpha=1.8$ and core radius $0.150$~Mpc.}
\label{fig:sims}
\end{figure}

We investigate the selection function for our galaxy sample by
simulating galaxy clusters of various richness and shapes at different
redshifts and examining their recovery fractions. Specifically we
generated clusters with random ellipticities uniformly between
$0.1<z<0.7$ using an $r^{-\alpha}$ profile for the galaxy distribution
\citep{Lubin-Postman-96} with $\alpha=1.8$ and $r_c=0.150$~Mpc and
using the luminosity function for red galaxies from
\cite{Brown-07}. These clusters are inserted 20 at a time in the
observed catalogs at random positions and redshifts, but avoiding the
locations where clusters were detected. In total we generate 10,000
simulated clusters with richness values, $N_{sim}$, of 15, 20, 25, 35,
50, 80 and 120 galaxies uniformly distributed in redshift.\footnote{These
richness values fold in the flux limit of the survey and the
membership prescription as described in the next section, so that they
are roughly comparable to the $N_{gal}$ values we give for the
detected clusters.}
We process each realization using the same
parameters as for the observed data and record the number of clusters
recovered as a function of redshift and galaxy richness. In
Figure~\ref{fig:sims} we show the results of this exercise where we
plot the recovery fraction as a function of redshift for the seven
cluster richnesses simulated. We conclude that for the rich clusters
$N_{sim}>50$ we are always nearly complete ($80-90\%$) for $z\le0.6$
while for poorer clusters we only detect at best $\sim30\%$ around
$z=0.3$.  

\section{Cluster Properties}

One of our main drivers in searching for clusters at optical
wavelengths is to correlate them with SZ detections in the new blind
SZ surveys. The signal to be detected in the mm-band experiments
(i.e., the $y$-distortion due to inverse Compton scattering) is
related to the number of hot electrons in the intracluster medium ,and
simulations have shown \citep[see][for
example]{Motl05,Nagai06,Sehgal07,Bhattacharya} that the SZE signal is
closely linked to cluster mass. Our analysis of the optical survey has
provided positions, redshifts, and fluxes of galaxies, from which we
infer the underlying cluster mass using scaling relations established
by the SDSS, where cluster masses were determined from weak lensing.

\begin{figure*}
\centerline{
\includegraphics[height=4.5in]{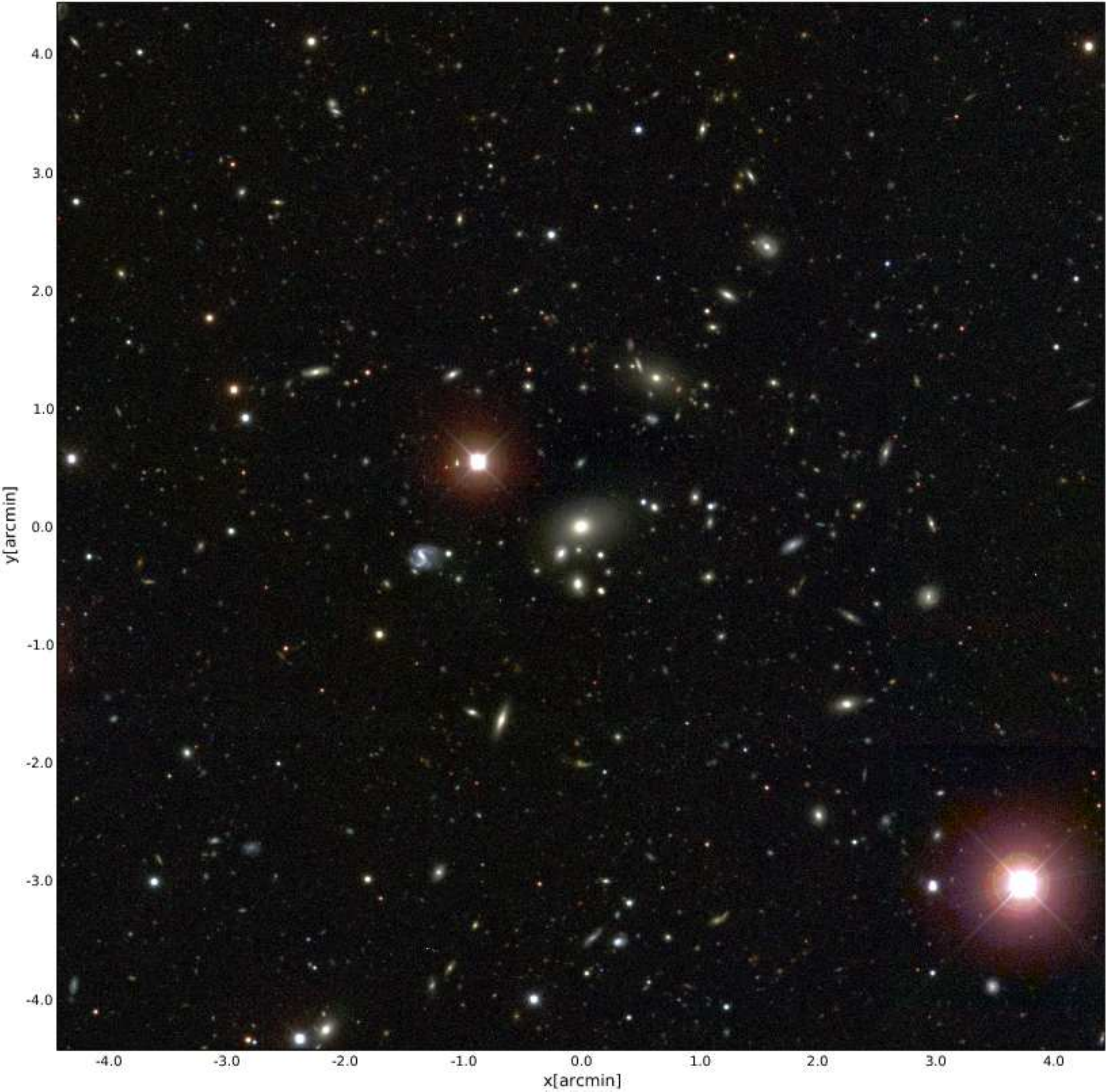}
\includegraphics[width=2.4in]{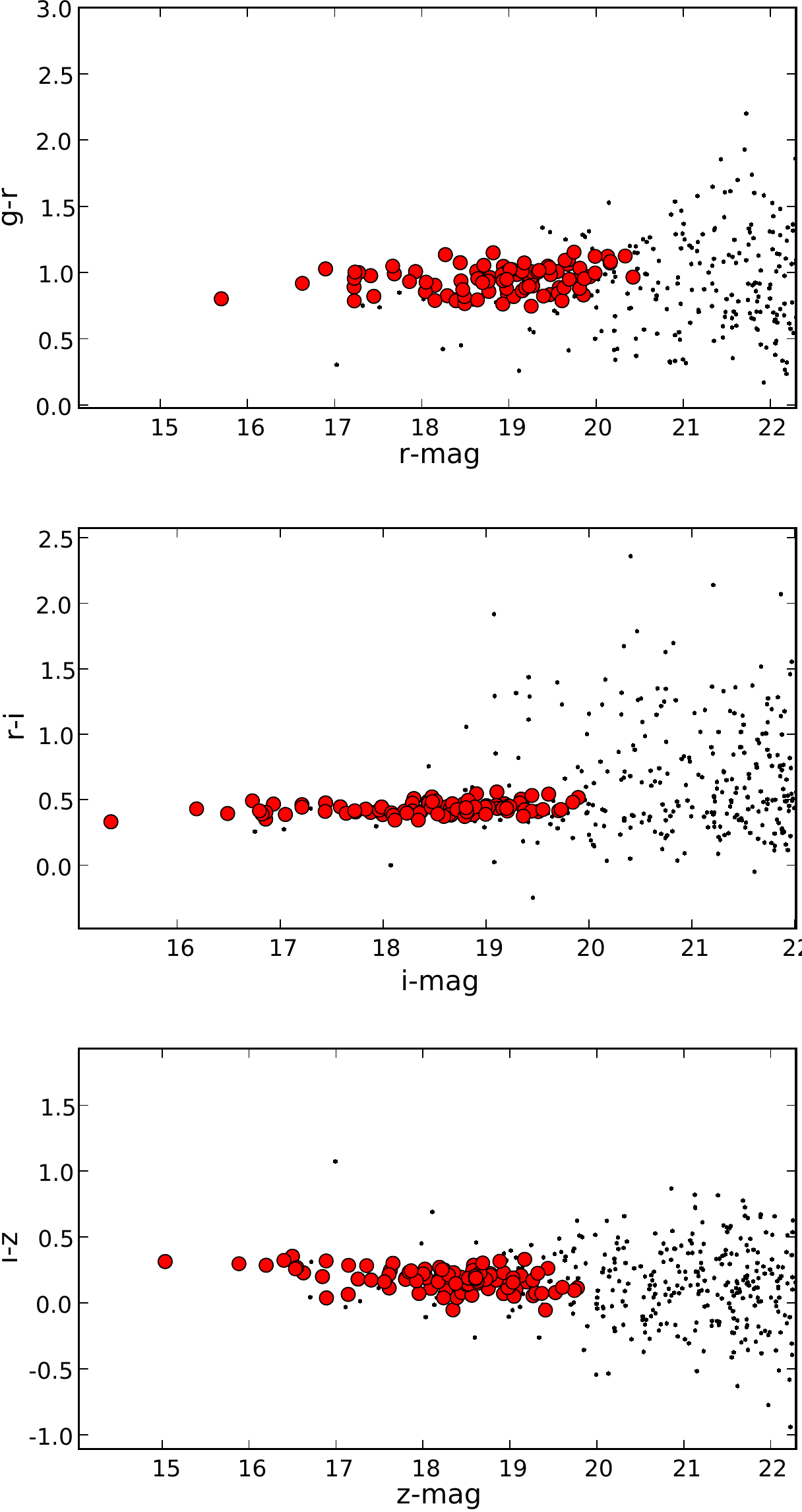}
}
\caption{The composed $gri$ color image (left panel) and
  color-magnitude relations (right) for X-ray cluster RXJ2325.6$-$5443
  from \cite{Mullis03}. Red points represent galaxies classified as
  E/S0 by BPZ that satisfy the conditions to be cluster members as
  described in the main text. Black dots are non-member galaxies in a
  $5'$ region near the cluster center.}
\label{fig:c0}
\end{figure*}

\subsection{Defining Cluster Membership}

The current state-of-the-art mass tracers for clusters of galaxies
using optically observed parameters \citep{Johnston07,Reyes08} have
been extracted from a sample of around $13,000$ optically-selected
clusters from the SDSS MaxBCG catalog \citep{MaxBCG}. In this paper,
we apply these scaling relations to our cluster sample and obtain mass
estimates from which we additionally predict SZ distortions. To be
fully consistent with previous analyses, we define membership and all
other relevant cluster observables following the same method as
\cite{Reyes08} and \cite{MaxBCG}.

We begin by examining each candidate-cluster peak in the density maps
and select the brightest elliptical galaxy in the cluster (BCG), which
is taken to be the initial center and redshift $z_o$ of the system. We
then use galaxies defined as E or E/S0s (i.e., BPZ SED types 0 and 1 only)
within a projected radius of $0.5h^{-1}$ Mpc
and redshift interval $|z-z_o| = |\Delta z|=0.05$ to obtain a local
color-magnitude relation (CMR) for each color combination, $g-r$,
$r-i$ and $i-z$, and the cluster mean redshift, $z_c$, for all cluster
members, using a $3\sigma$ median sigma-clipping algorithm.  We use
these to determine $N_{\rm 1Mpc}$, the number of galaxies within
$1h^{-1}$Mpc of the cluster center. Formally, we compute
$N_{gal}=N_{\rm 1Mpc}$ by including those galaxies within a projected
1$h^{-1}$Mpc from the cluster center that satisfy three conditions:
(a) the galaxy must have the SED of an E, E/S0 according to BPZ, (b)
it must have the appropriate color to be a cluster member (i.e.,
colors within $3\sigma$ of the local CMR for all color combinations)
and (c) it must have the right luminosity (i.e., dimmer than the BCG
and brighter than $0.4L^*$, where we use the corresponding absolute
magnitude $M_i^*$ from \cite{Brown-07} redshifted with the elliptical
SED template from BPZ). We designated cluster members according to the
estimated cluster size $R_{200}$, defined as the radius at which the
cluster galaxy density is $200\Omega_m^{-1}$ times the mean space
density of galaxies in the present Universe. We estimated the scaled
radius $R_{200}$ using the empirical relation from \cite{Hansen-05},
$R_{200}= 0.156N_{\rm 1Mpc}^{0.6} h^{-1}$Mpc which is derived from the
SDSS and we assume it holds beyond $z\sim0.3$ for our higher redshift clusters.

In our analysis we use \n200, \L200, and \LBCG\  to scale cluster
optical parameters with mass, following \cite{Reyes08}.
The cluster richness, \n200, is the number of E/S0 galaxies within
$R_{200}$ with colors and luminosities that satisfy conditions (b) and
(c) above. Similarly, $L_{200}$ is the total rest-frame integrated
$r$-band luminosity of all member galaxies included in \n200 in units
of $10^{10}h^{-2}L_{\sun}$ and $L_{\rm BCG}$ is the rest-frame
$r$-band luminosity of the BCG.

In order to have reliable estimates of $N_{\rm gal}$ it is necessary
to determine the galaxy background contamination and implement an
appropriate background subtraction method. The lack of spectroscopic
redshifts in our sample only allowed a statistical removal of
unrelated field galaxies with similar colors and redshifts that were
projected along the line of sight to each cluster. We assumed that the
presence of a cluster at some redshift is independent of the field
population seen in projection.  Therefore we estimate the surface
number density of ellipticals in an annulus surrounding the cluster
(within $R_{200}<r<2 R_{200}$) with $\Delta z=0.05$ and the same
colors as the cluster members. We measure this background contribution
around the outskirts of each cluster and obtain a corrected value
$N_{gal}$ which is used to compute $R_{200}$ and then corresponding
values of \n200 and $L_{200}$. The magnitude of the correction ranges
between $15-20$\%. We will refer to the corrected values hereafter.
Moreover as our analysis is based on a magnitude-limited sample it is
worth considering the fraction of lower luminosity galaxies that will
fall below our magnitude limit ($i=22.5$) at higher redshifts. As in
Menanteau \& Hughes (2009), if we make the assumption that the cluster
population is like that of the five clusters at $z<0.2$ in our sample
(see Table~\ref{tab:data2}) and $M^*$ evolves passively, then we can compute the
fraction of $L_{200}$ missed for clusters at higher redshifts. We
estimate that we are missing 4\%, 13\%, 29\% and 38\% of the cluster
luminosity at $z=0.4, 0.5, 0.6$ and $0.7$ respectively.  Given the
uncertainty in this correction factor, we do not include it in our
quoted luminosity values for the higher redshift clusters.  This means
our cluster masses, $M(L_{200}$), are underestimated by roughly these
factors.

\begin{figure*}
\centerline{
\includegraphics[width=2.1in]{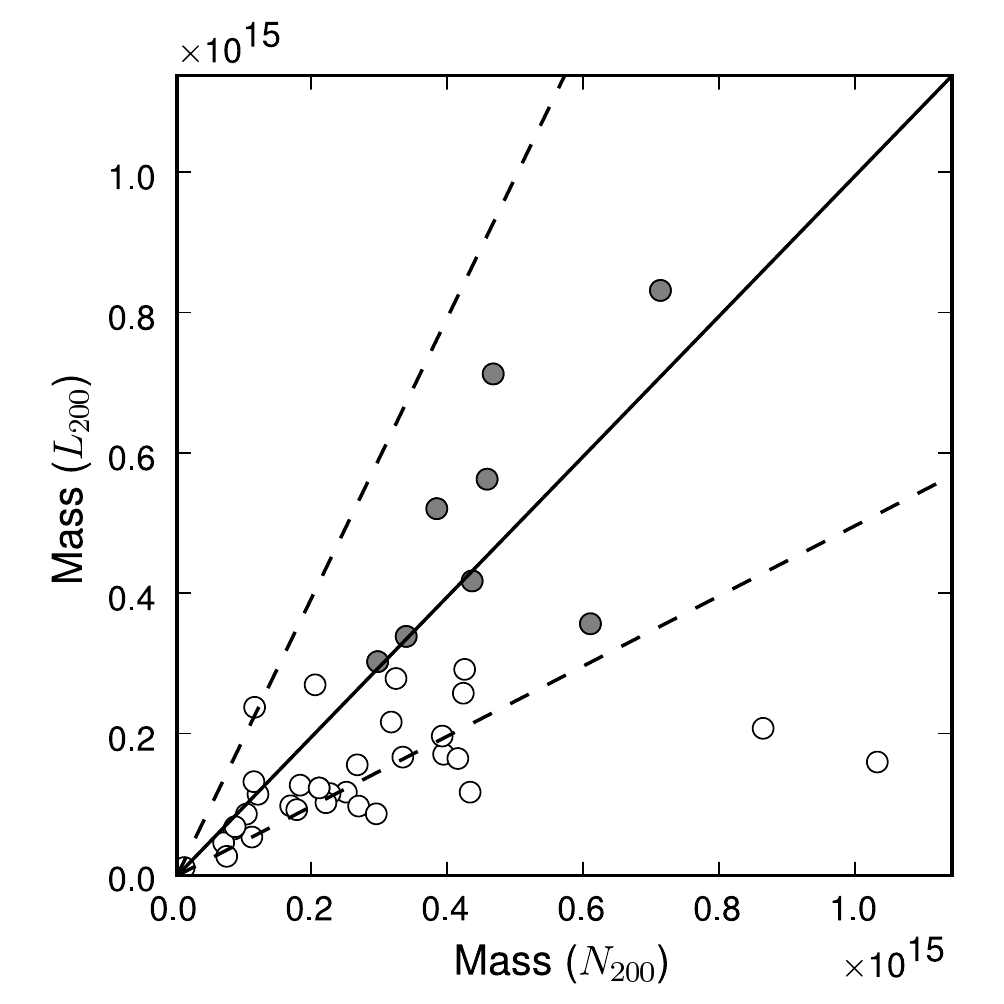}
\includegraphics[width=2.1in]{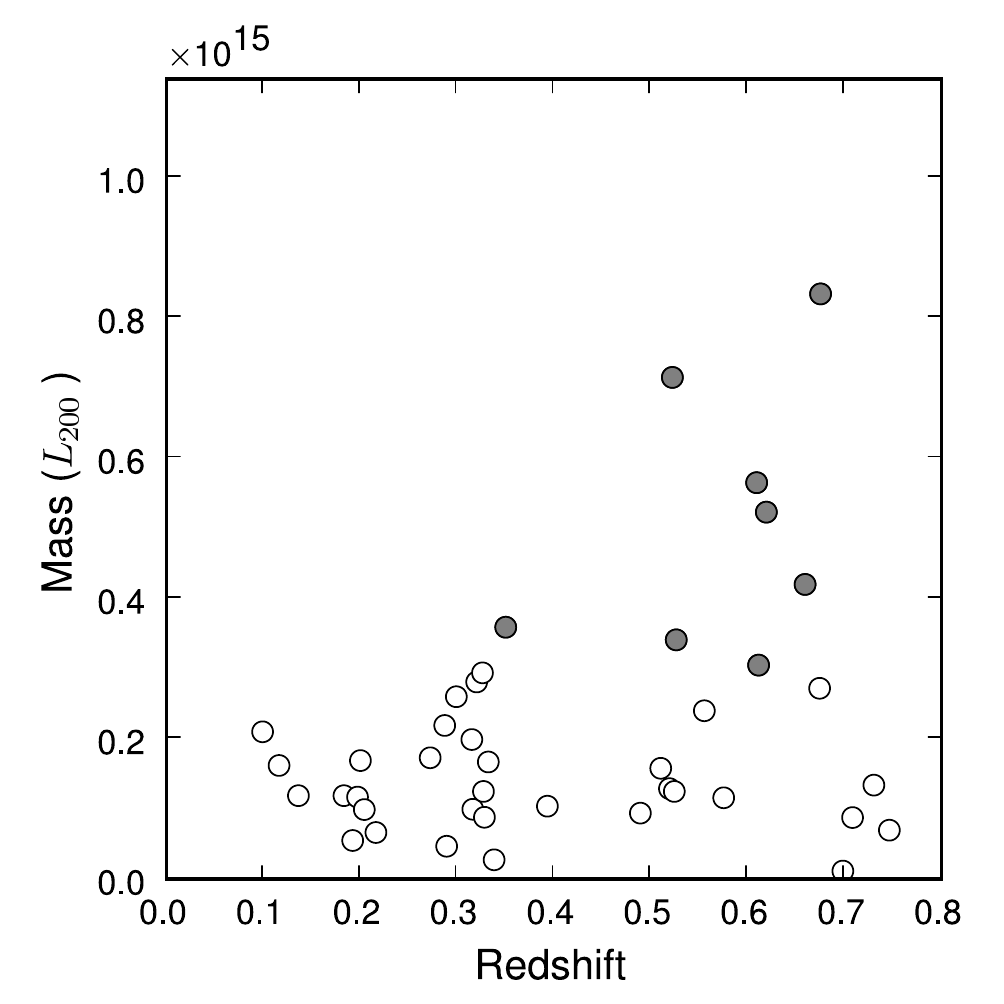}
\includegraphics[width=2.1in]{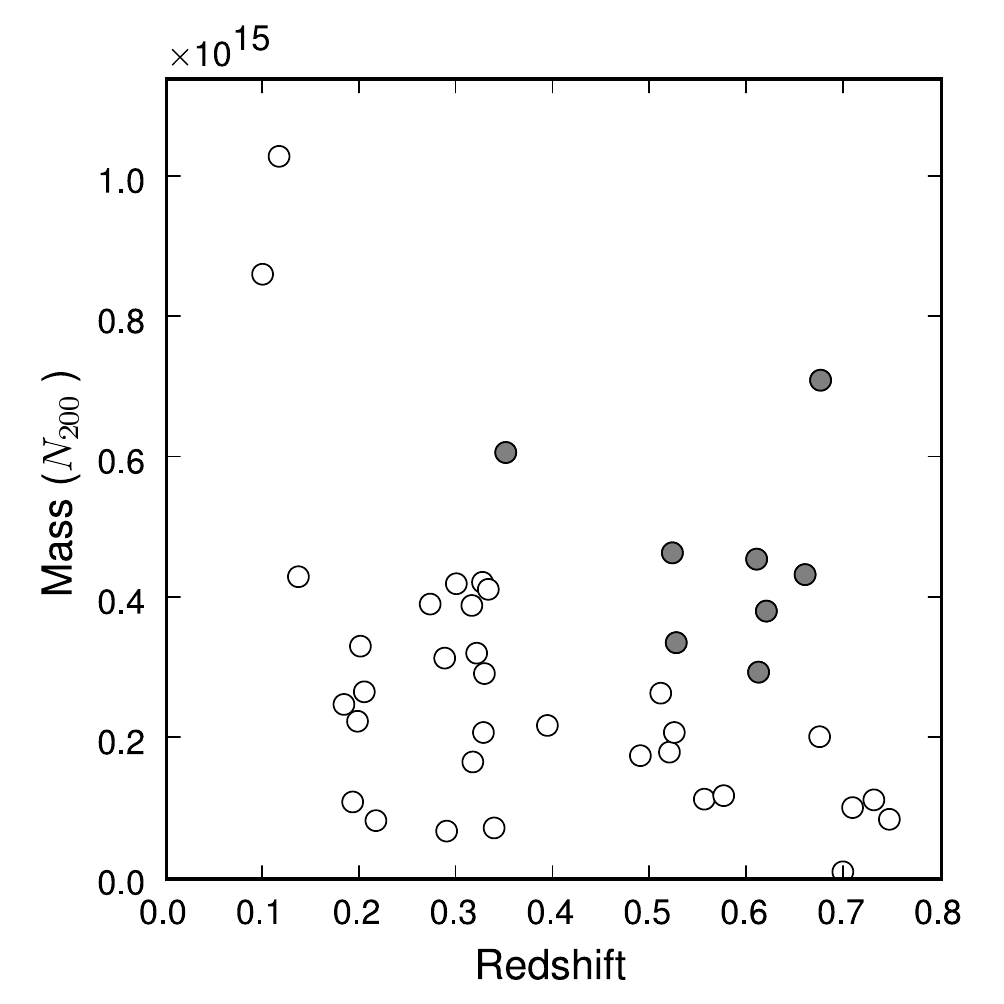}
}
\caption{The mass observable parameters for the 38 clusters in our
  sample. Filled circles represent the eight new clusters above the mass limit
  $M(L_{200}) > 3\times 10^{14}~M_{\sun}$. The left panel shows the relation
  between the mass predictions using $L_{200}$ vs. \n200 as the main
  parameter; the agreement is generally to within a factor of two,
  except for the nearest systems. The central and right panel show the
  mass estimates as a function of the cluster redshift using $L_{200}$
  and \n200 respectively, in units of solar mass.}
\label{fig:3M}
\end{figure*}

\subsection{Recovery of Known Clusters}

The area covered by the BCS in the 23hr region is a
virtual desert in terms of known clusters and spectroscopic
redshifts for galaxies with $z>0.1$. We found one catalogued X-ray
selected cluster from the 160~deg$^2$ {\em ROSAT} survey
\citep{Mullis03}: RXJ2325.6$-$5443 at $z=0.102$ with an
X-ray flux of $F_X=2.2\times 10^{-13}$~erg cm$^{-2}$ s$^{-1}$ in the
0.5--2 keV band. Our cluster finding algorithm easily recovered this
cluster and produced a photometric redshift estimate of $\langle z
\rangle = 0.10 \pm 0.02$.  Figure~\ref{fig:c0} shows the $gri$ color
composite optical image of the cluster as well as the color magnitude
diagrams for cluster members. Using the techniques described below we
estimate the mass of RXJ2325.6$-$5443 to be $M(L_{200})=2.1\times
10^{14}$ which is just below the detectability limit of ACT and
therefore will not be included in our SZE predictions. From the
$M$-$T_X$ \citep{Evrard96} and $L_X$-$T_X$ \citep{ArnaudEvrard99} relations
we estimate a temperature of $kT\sim 2.2$ keV and an X-ray flux of
$F_X \sim 5\times10^{-13}$ erg cm$^{-2}$ s$^{-1}$ (0.5--2 keV band)
which is in rough agreement (factor of 2) with the published value.
Considering the entire sample of 38 clusters, RXJ2325.6$-$5443 is the
closest and, based on our estimated masses from the optical properties
and the $M$-$T_X$ and $L_X$-$T_X$ relations, has the highest predicted X-ray
flux of the sample.  Still, even this value is below the X-ray
detection threshold of the {\em ROSAT} All Sky Survey (RASS) and so,
as expected, we find that none of our new optical clusters are
significant RASS X-ray sources.

\subsection{Cluster Mass Estimation}

\begin{figure}
\includegraphics[width=\columnwidth]{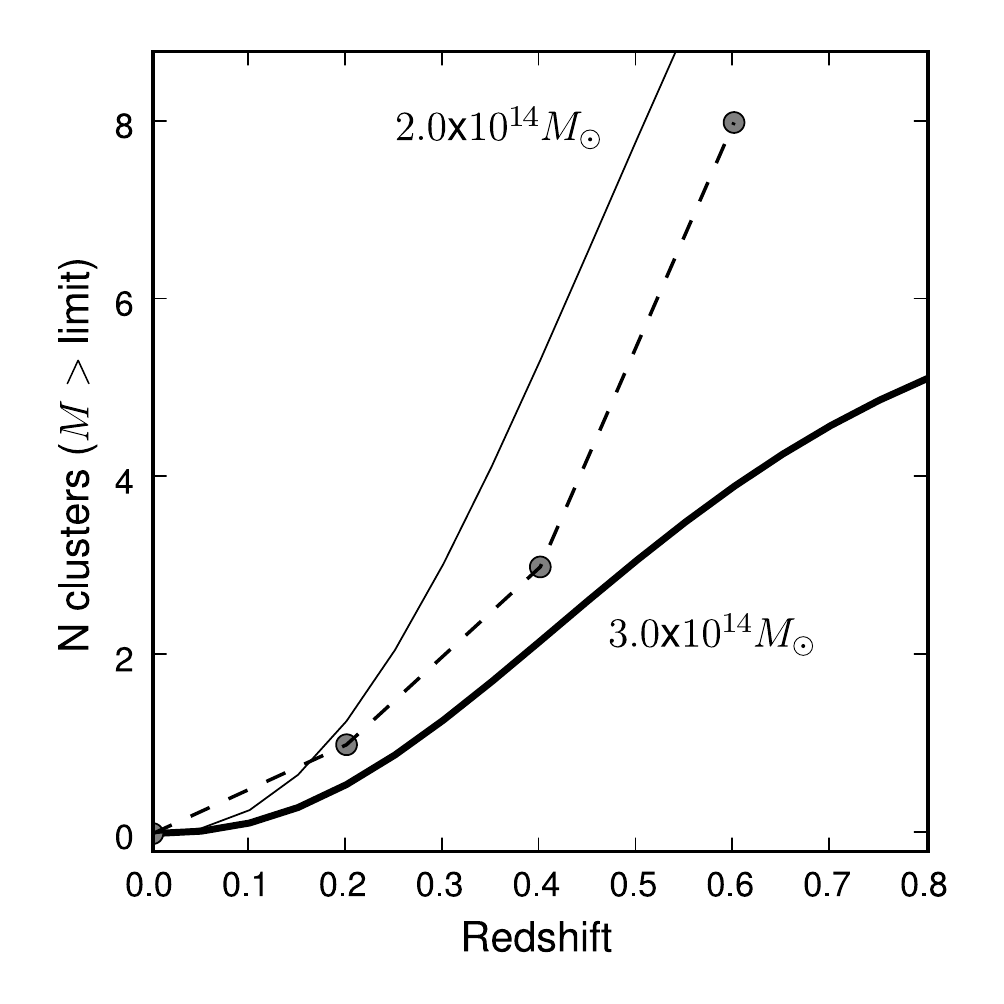}
\caption{The gray circles show the total number of clusters with
  $M(L_{200}) >3\times 10^{14}\,M_\odot$ observed in the $\simeq8$~deg$^2$
  region analyzed here. The thin and thick lines show the predicted number of
  clusters in the region for cluster masses greater than $2\times
  10^{14}\, M_\odot$ and $3\times 10^{14}\, M_\odot$ respectively,
  using the WMAP 5-year best-fit cosmological model and the
  \cite{Jenkins01} halo mass function.}
\label{fig:Nclusters}
\end{figure}

\begin{deluxetable*}{rrrrrrrrrrr}
\tablecaption{Mass-richness power-law function Best fitting parameters}
\tablewidth{0pt}
\tablehead{
\multicolumn{1}{r}{} & \multicolumn{2}{r}{$(10^{10}h^{-2}L_{\sun})$} &\multicolumn{8}{r}{}\\
\colhead{Redshift} & 
\colhead{$a_N$} & 
\colhead{$a_L$} & 
\colhead{$b_N$} & 
\colhead{$b_L$} &
\colhead{$M_N^0$} &
\colhead{$\alpha_N$} &
\colhead{$\gamma_N$} &
\colhead{$M_L^0$} &
\colhead{$\alpha_L$} &
\colhead{$\gamma_L$} 
}
\startdata
$0.10<z<0.23$ & 1.54   & 7.77  & 0.41 & 0.67 & $1.27\pm0.08$ & $1.20\pm0.09$ & $0.71\pm 0.14$ & $1.81\pm0.15$ & $1.27\pm0.17$ & $0.40\pm0.23$ \\ 
$0.23<z<0.70$ & 1.64   & 7.92  & 0.43 & 0.66 & $1.57\pm0.14$ & $1.12\pm0.15$ & $0.34\pm 0.24$ & $1.76\pm0.22$ & $1.30\pm0.29$ & $0.26\pm0.41$ \\ 
\enddata
\label{tab:pars}
\end{deluxetable*}

In this section we use the mass-richness relations based on \N1Mpc\ to
weigh our new optical clusters. Both \cite{Johnston07}
and \cite{Reyes08} found that the luminosity-mass and richness-mass
relations were well described by power-law functions and they measured
the normalizations and slopes in these relations using $\chi^2$
minimization. Their values are in broad agreement, but we will use the
fits provided by \cite{Reyes08} since they restrict their fits to 
clusters with
$N_{200} > 10$ and give results for two redshift bins: $0.10
<z<0.23$ and $0.23<z<0.30$. It is important to note that our
clusters go well beyond $z=0.3$ and that we extrapolate the relation
in the last redshift bin for clusters with $z>0.3$. 
We investigated the two fitting functions based on $L_{200}$ and
\n200, (see section 5.2.1 from \citealt{Reyes08} for full details),
which are described as:
\begin{equation}
M(N_{200},L_{\rm BCG}) = M_N^0(N_{200}/20)^{\alpha_N}(L_{\rm BCG}/\bar{L}_{\rm BCG}^{(N)})^{\gamma_N} 
\label{eq:M1}
\end{equation}
\begin{equation}
M(L_{200},L_{\rm BCG}) = M_L^0(L_{200}/40)^{\alpha_L}(L_{\rm BCG}/\bar{L}_{\rm BCG}^{(L)})^{\gamma_L}
\label{eq:M2}
\end{equation}
where $M$ is the mass observational equivalent of
$M_{200\bar{\rho}}$\footnote{$M_{200\bar{\rho}}$ is the halo mass
  enclosed within $R_{200}$, defined as a radius of spherical volume
  within which the mean density is 200 times the critical density.} in
units of $10^{14}M_{\sun}$, $L_{200}$ is in units of
$10^{10}h^{-2}L_{\sun}$ and the \LBCG\ dependence is normalized by its
mean value. This is also described by a power-law function for a given
value of \L200\ and \n200:
\begin{equation}
\bar{L}_{\rm BCG}^{(N)} \equiv \bar{L}_{\rm BCG}(N_{200}) = a_N N_{200}^{b_N}
\label{eq:L1}
\end{equation}
\begin{equation}
\bar{L}_{\rm BCG}^{(L)} \equiv \bar{L}_{\rm BCG}(L_{200}) = a_L L_{200}^{b_L}
\label{eq:L2}
\end{equation}
The published best-fitting parameters for $M^0$, $\alpha$ and $\gamma$
in Eqs.~(\ref{eq:M1}) and (\ref{eq:M2}) as well as the values of $a$, $b$ for Eqs.~(\ref{eq:L1}) and
(\ref{eq:L2}) are shown in Table~\ref{tab:pars}. The combination of equations
(\ref{eq:M1}), (\ref{eq:M2}) and (\ref{eq:L1}), (\ref{eq:L2}) for \n200 and \L200 respectively enable us to
obtain mass estimates for any cluster with $N_{\rm gal}>10$.

The left panel of Fig.~\ref{fig:3M} compares the masses obtained using
$N_{200}$ and $L_{200}$ for our 38 optical clusters.  The solid
line denotes equality between the estimates while the two dashed lines
show a factor of 2 range.  Most clusters fall within this range,
establishing a lower bound on our mass error. The two most significant
outliers correspond to nearby clusters (see the right panel of
Fig.~\ref{fig:3M}) for which $N_{200}$ is evidently overpredicting the
mass compared to $L_{200}$.  In one of these cases (which is
RXJ2325.6$-$5443), the mass inferred by $N_{200}$ grossly overpredicts
(by more than an order of magnitude) the estimated X-ray flux. The
natural concentration of luminosities in clusters (i.e., that more
luminous galaxies dominate the central regions of clusters) suggests
that the mass estimate derived from $L_{200}$ should be more robust
than that from $N_{200}$ as a function of redshift.  For these reasons
we use the cluster mass estimate derived from $L_{200}$ in predicting
the SZE signal. Hereafter we refer to this mass as $M_{200}^L$.

Table~\ref{tab:data} displays the 8 clusters with $M(L_{200}) > 3\times 10^{14}
M_\odot$ in the 8 square-degree sky area covered by our
analysis. These clusters also have  $M(N_{200})$ above the same mass limit.
In Figure~\ref{fig:Nclusters} we show the cumulative number distribution
of these clusters as a function of redshift, compared with expectations
from simulations (H.~Trac, private communication), using the mass function
of dark matter haloes from \cite{Jenkins01} and WMAP5 cosmology
\citep{WMAP5}.  The heavy solid curve shows expectations for a mass
limit of $3\times 10^{14} M_{\sun}$, while the light solid curve
indicates a mass limit of $2\times 10^{14} M_{\sun}$.  This figure
demonstrates that the number of clusters we observe is consistent with
a mass limit in the range of $2-3\times 10^{14} M_{\sun}$. 
In Table~\ref{tab:data2} we display the properties for the remaining clusters
with $M(L_{200}) < 3\times 10^{14} M_\odot$.

\section{Predictions of the Sunyaev-Zel'dovich Effect Signal}

Scaling relations between the integrated thermal SZE signal and
cluster mass have emerged from current N-body plus hydrodynamic
simulations of galaxy clusters. In this section we use these relations
and our optically-derived mass estimates, $M_{200}^{L}$, to predict
the SZE signal to be observed by ACT and SPT when these experiments
survey this sky region.

The SZE signal consists of small distortions to the CMB spectrum
originating from inverse Compton scattering by electrons in the hot
plasma of clusters of galaxies \citep{SZ80}. Here we consider the
thermal SZ flux $Y$, defined as the integrated Compton $y-$parameter,
\begin{equation}
Y = d_A^2(z) \int_{\Omega} y d\Omega = \frac{k_{\rm B} \sigma_{\rm T}}{m_ec^2}\int_Vn_eT_edV,
\end{equation}
where $n_e$ and $T_e$ are the number density and temperature of hot
electrons in the cluster, $m_e$ is electron rest mass, $c$ is the
speed of light, $\sigma_{\rm T}$ is the Thompson scattering cross
section and $d_A(z)$ is the angular diameter distance. The projected
area, $dA$, and solid angle, $d\Omega$, are related by the angular
diameter distance as $dA = d^2_A(z) d\Omega$.

Self-similar scaling relations \citep{Kaiser86} predict that the
virialized mass $M$ in clusters scales with the gas temperature as
$M\propto T^{3/2}/E(z)$ where $E(z) = (\Omega_m(1+z)^3 +
\Omega_{\Lambda})^{1/2}$ for a flat cosmology. If clusters were
isothermal, we would expect their SZE signal to scale like $Y\propto
f_{\rm gas} M_{\rm halo}T$ and therefore the self-similar SZ flux-mass
scaling relation should have the shape $Y\propto f_{\rm gas} M^{5/3}
E^{2/3}(z)$ where $f_{\rm gas}$ is the cluster mass fraction. However,
clusters are not always isothermal or in hydrostatic equilibrium and
physical processes like star formation and feedback will also
contribute to deviations from self-similarity.

Prompted by the upcoming SZ surveys, several studies have
characterized in detail the $Y-M$ scaling relation using simple
power-law fits to cosmological N-body simulations. Here we will use
the recent $Y-M$ fits from \cite{Sehgal07}, who included gas
simulations employing small-scale cluster physics such as star
formation and feedback into a large cosmological N-body simulation.
>From their catalog of $\sim10^5$ simulated clusters with
$M_{200}>7.5\times 10^{13}M_{\sun}$, \cite{Sehgal07} fit the relation 
\begin{equation}
\frac{Y_{200}}{E(z)^{2/3}} =  10^{\beta} \left(\frac{M_{200}}{10^{14}  M_{\sun}}\right) ^{\alpha},
\label{eq:y200}
\end{equation}
where $Y_{200}$ and $M_{200}$ are the projected SZ Compton
$y$-parameter and mass respectively in a disk of radius $R_{200}$. 
We use the best fit values $\alpha=1.876\pm0.005$ and
$\beta=-5.4774\pm0.0009$ for all clusters regardless of redshift, as
the redshift dependence of the fits is very weak \cite[see Table~2
  from][]{Sehgal07}. The power-law index that they report is slightly
steeper than ones quoted by some previous hydrodynamic simulations
\citep[see][for example]{Hernandez06, Nagai06, Motl05}, which
\cite{Sehgal07} attribute to their more realistic feedback
prescription (i.e., including the effects of AGN and supernovae). We
use the power-law model from equation~\ref{eq:y200} and the optical
$M_{200}^L$ mass estimates for our 8 new massive clusters to predict
the integrated $Y_{200}$ signal to be observed by ACT and
SPT. Table~\ref{tab:data} gives the results in physical units of
Mpc$^{2}$ and observable ones of arcmin$^2$.

\begin{deluxetable*}{rrrrrrrrr}
\tablecaption{Optical Clusters with $M(L_{200})>3\times 10^{14}M_{\sun}$}
\tablewidth{0pt}
\tablehead{
\multicolumn{5}{c}{} & \multicolumn{2}{c}{$[M_{\sun}]$} & \multicolumn{1}{c}{[Mpc$^2$]} & \multicolumn{1}{c}{[arcmin$^2$]} \\
\colhead{ID} & 
\colhead{$z$} & 
\colhead{$N_{\rm gal}$} &
\colhead{\n200} &
\colhead{$L_{200}[L_{\sun}]$} &
\colhead{$M(N_{200})$} &
\colhead{$M(L_{200})$} &
\colhead{$Y_{200}/E(z)^{2/3}$} &
\colhead{$Y_{200}$} 
}
\startdata
SCSO~J233430.2--543647.5 & 0.35 & $ 32.1\pm  5.8$ & $ 43.8\pm  7.0$ & $ 1.8 \times 10^{12}\pm2.9 \times 10^{10}$ & $6.1 \times 10^{14}$ & $3.6 \times 10^{14}$ & $3.5 \times 10^{-05}$ & $4.8 \times 10^{-04}$ \\
SCSO~J233556.8--560602.3 & 0.52 & $ 31.4\pm  5.9$ & $ 33.5\pm  6.4$ & $ 3.3 \times 10^{12}\pm1.4 \times 10^{11}$ & $4.6 \times 10^{14}$ & $7.2 \times 10^{14}$ & $1.2 \times 10^{-04}$ & $1.1 \times 10^{-03}$ \\
SCSO~J233425.6--542718.0 & 0.53 & $ 26.1\pm  5.4$ & $ 26.9\pm  5.6$ & $ 1.8 \times 10^{12}\pm9.5 \times 10^{10}$ & $3.4 \times 10^{14}$ & $3.4 \times 10^{14}$ & $3.2 \times 10^{-05}$ & $3.0 \times 10^{-04}$ \\
SCSO~J232211.0--561847.4 & 0.61 & $ 31.1\pm  5.7$ & $ 34.1\pm  6.1$ & $ 2.8 \times 10^{12}\pm1.1 \times 10^{11}$ & $4.6 \times 10^{14}$ & $5.6 \times 10^{14}$ & $7.9 \times 10^{-05}$ & $6.8 \times 10^{-04}$ \\
SCSO~J233731.7--560427.9 & 0.61 & $ 25.6\pm  5.4$ & $ 23.1\pm  5.3$ & $ 1.7 \times 10^{12}\pm1.7 \times 10^{11}$ & $3.0 \times 10^{14}$ & $3.0 \times 10^{14}$ & $2.6 \times 10^{-05}$ & $2.2 \times 10^{-04}$ \\
SCSO~J234012.6--541907.2 & 0.62 & $ 22.7\pm  5.2$ & $ 24.9\pm  5.4$ & $ 2.4 \times 10^{12}\pm1.4 \times 10^{11}$ & $3.8 \times 10^{14}$ & $5.2 \times 10^{14}$ & $6.9 \times 10^{-05}$ & $5.8 \times 10^{-04}$ \\
SCSO~J234004.9--544444.8 & 0.66 & $ 30.7\pm  5.8$ & $ 37.5\pm  6.7$ & $ 2.3 \times 10^{12}\pm2.4 \times 10^{11}$ & $4.3 \times 10^{14}$ & $4.2 \times 10^{14}$ & $4.6 \times 10^{-05}$ & $3.8 \times 10^{-04}$ \\
SCSO~J232829.7--544255.4 & 0.68 & $ 74.6\pm  8.7$ & $ 65.7\pm  9.7$ & $ 4.4 \times 10^{12}\pm4.3 \times 10^{11}$ & $7.1 \times 10^{14}$ & $8.3 \times 10^{14}$ & $1.6 \times 10^{-04}$ & $1.3 \times 10^{-03}$ \\
\enddata
\label{tab:data}
\tablecomments{Catalog of the optical clusters with mass estimates
  $>3\times 10^{14} M_{\sun}$ from the $M(L_{200})$ values. Each cluster's 
  redshift is the mean photometric redshift computed using the elliptical 
  in the center of the cluster. The ID is based on the position of the
  BCG.}
\end{deluxetable*}

\begin{deluxetable*}{rrrrrrr}
\tablecaption{Optical Clusters with $M(L_{200})<3\times 10^{14}M_{\sun}$}
\tablewidth{0pt}
\tablehead{
\multicolumn{5}{c}{} & \multicolumn{2}{c}{$[M_{\sun}]$}   \\
\colhead{ID} & 
\colhead{$z$} & 
\colhead{$N_{\rm gal}$} &
\colhead{\n200} &
\colhead{$L_{200}[L_{\sun}]$} &
\colhead{$M(N_{200})$} &
\colhead{$M(L_{200})$} 
}
\startdata
SCSO~J232540.2--544430.9 & 0.10 & $123.9\pm 11.2$ & $278.1\pm 18.5$ & $ 3.3 \times 10^{12}\pm2.8 \times 10^{10}$ & $8.6 \times 10^{14}$ & $2.1 \times 10^{14}$  \\
SCSO~J232230.9--541608.3 & 0.12 & $ 69.5\pm  8.4$ & $145.3\pm 12.8$ & $ 1.7 \times 10^{12}\pm1.1 \times 10^{10}$ & $1.0 \times 10^{15}$ & $1.6 \times 10^{14}$  \\
SCSO~J233000.4--543707.7 & 0.14 & $ 38.0\pm  6.4$ & $ 39.0\pm  7.2$ & $ 1.0 \times 10^{12}\pm1.0 \times 10^{10}$ & $4.3 \times 10^{14}$ & $1.2 \times 10^{14}$  \\
SCSO~J232419.6--552548.9 & 0.18 & $ 39.3\pm  6.5$ & $ 36.9\pm  7.0$ & $ 1.4 \times 10^{12}\pm1.8 \times 10^{10}$ & $2.5 \times 10^{14}$ & $1.2 \times 10^{14}$  \\
SCSO~J233106.9--555119.5 & 0.19 & $ 26.3\pm  5.5$ & $ 35.3\pm  6.4$ & $ 9.7 \times 10^{11}\pm2.9 \times 10^{10}$ & $1.1 \times 10^{14}$ & $5.5 \times 10^{13}$  \\
SCSO~J233252.9--561454.1 & 0.20 & $ 34.0\pm  6.1$ & $ 43.4\pm  7.2$ & $ 1.5 \times 10^{12}\pm2.2 \times 10^{10}$ & $2.2 \times 10^{14}$ & $1.2 \times 10^{14}$  \\
SCSO~J233215.5--544211.6 & 0.20 & $ 43.5\pm  6.8$ & $ 42.6\pm  7.6$ & $ 1.8 \times 10^{12}\pm3.4 \times 10^{10}$ & $3.3 \times 10^{14}$ & $1.7 \times 10^{14}$  \\
SCSO~J233037.1--554338.8 & 0.20 & $ 27.5\pm  5.6$ & $ 35.2\pm  6.4$ & $ 1.1 \times 10^{12}\pm1.7 \times 10^{10}$ & $2.7 \times 10^{14}$ & $9.9 \times 10^{13}$  \\
SCSO~J233550.6--552820.4 & 0.22 & $ 14.5\pm  4.4$ & $ 10.6\pm  3.5$ & $ 7.4 \times 10^{11}\pm2.2 \times 10^{10}$ & $8.3 \times 10^{13}$ & $6.6 \times 10^{13}$  \\
SCSO~J232200.4--544459.7 & 0.27 & $ 34.6\pm  6.0$ & $ 41.0\pm  6.9$ & $ 1.2 \times 10^{12}\pm1.8 \times 10^{10}$ & $3.9 \times 10^{14}$ & $1.7 \times 10^{14}$  \\
SCSO~J233522.6--553237.0 & 0.29 & $ 31.4\pm  5.9$ & $ 32.1\pm  6.3$ & $ 1.5 \times 10^{12}\pm2.4 \times 10^{10}$ & $3.2 \times 10^{14}$ & $2.2 \times 10^{14}$  \\
SCSO~J233807.5--560304.9 & 0.30 & $ 32.0\pm  5.9$ & $ 37.7\pm  6.6$ & $ 1.6 \times 10^{12}\pm3.2 \times 10^{10}$ & $4.2 \times 10^{14}$ & $2.6 \times 10^{14}$  \\
SCSO~J232956.0--560808.3 & 0.32 & $ 39.6\pm  6.5$ & $ 37.0\pm  6.7$ & $ 1.3 \times 10^{12}\pm2.3 \times 10^{10}$ & $3.9 \times 10^{14}$ & $2.0 \times 10^{14}$  \\
SCSO~J232839.5--551353.8 & 0.32 & $ 40.3\pm  6.5$ & $ 18.9\pm  5.2$ & $ 7.9 \times 10^{11}\pm2.2 \times 10^{10}$ & $1.7 \times 10^{14}$ & $1.0 \times 10^{14}$  \\
SCSO~J232633.6--550111.5 & 0.32 & $ 74.3\pm  8.7$ & $ 35.2\pm  7.4$ & $ 1.9 \times 10^{12}\pm4.5 \times 10^{10}$ & $3.2 \times 10^{14}$ & $2.8 \times 10^{14}$  \\
SCSO~J233753.8--561147.6 & 0.33 & $ 33.2\pm  5.9$ & $ 41.3\pm  6.7$ & $ 1.8 \times 10^{12}\pm3.3 \times 10^{10}$ & $4.2 \times 10^{14}$ & $2.9 \times 10^{14}$  \\
SCSO~J232156.4--541428.8 & 0.33 & $ 20.1\pm  4.8$ & $ 19.9\pm  4.7$ & $ 8.6 \times 10^{11}\pm1.0 \times 10^{10}$ & $2.1 \times 10^{14}$ & $1.2 \times 10^{14}$  \\
SCSO~J233003.6--541426.7 & 0.33 & $ 29.6\pm  5.7$ & $ 30.4\pm  5.9$ & $ 6.6 \times 10^{11}\pm1.8 \times 10^{10}$ & $2.9 \times 10^{14}$ & $8.8 \times 10^{13}$  \\
SCSO~J233231.4--540135.8 & 0.33 & $ 45.9\pm  6.9$ & $ 42.7\pm  7.1$ & $ 1.2 \times 10^{12}\pm2.2 \times 10^{10}$ & $4.1 \times 10^{14}$ & $1.7 \times 10^{14}$  \\
SCSO~J233110.6--555213.5 & 0.39 & $ 21.1\pm  4.9$ & $ 20.8\pm  4.9$ & $ 7.3 \times 10^{11}\pm1.7 \times 10^{10}$ & $2.2 \times 10^{14}$ & $1.0 \times 10^{14}$  \\
SCSO~J233618.3--555440.3 & 0.49 & $ 17.4\pm  4.6$ & $ 15.3\pm  4.2$ & $ 6.3 \times 10^{11}\pm3.6 \times 10^{10}$ & $1.8 \times 10^{14}$ & $9.4 \times 10^{13}$  \\
SCSO~J233706.3--541903.8 & 0.51 & $ 25.5\pm  5.3$ & $ 29.9\pm  5.9$ & $ 1.2 \times 10^{12}\pm8.5 \times 10^{10}$ & $2.6 \times 10^{14}$ & $1.6 \times 10^{14}$  \\
SCSO~J233816.9--555331.1 & 0.52 & $ 19.8\pm  4.7$ & $ 19.8\pm  4.7$ & $ 9.7 \times 10^{11}\pm3.4 \times 10^{10}$ & $1.8 \times 10^{14}$ & $1.3 \times 10^{14}$  \\
SCSO~J232619.8--552308.8 & 0.52 & $ 18.8\pm  4.7$ & $ 18.2\pm  4.5$ & $ 8.1 \times 10^{11}\pm6.1 \times 10^{10}$ & $2.1 \times 10^{14}$ & $1.2 \times 10^{14}$  \\
SCSO~J232215.9--555045.6 & 0.56 & $ 11.0\pm  4.0$ & $  7.4\pm  2.9$ & $ 1.2 \times 10^{12}\pm6.1 \times 10^{10}$ & $1.1 \times 10^{14}$ & $2.4 \times 10^{14}$  \\
SCSO~J232247.6--541110.1 & 0.57 & $ 14.9\pm  4.3$ & $ 11.5\pm  3.6$ & $ 8.2 \times 10^{11}\pm4.3 \times 10^{10}$ & $1.2 \times 10^{14}$ & $1.2 \times 10^{14}$  \\
SCSO~J232342.3--551915.1 & 0.67 & $ 18.1\pm  4.7$ & $ 18.5\pm  4.6$ & $ 1.7 \times 10^{12}\pm1.5 \times 10^{11}$ & $2.0 \times 10^{14}$ & $2.7 \times 10^{14}$  \\
SCSO~J233403.7--555250.7 & 0.71 & $ 11.5\pm  4.1$ & $ 10.0\pm  3.3$ & $ 6.5 \times 10^{11}\pm1.5 \times 10^{11}$ & $1.0 \times 10^{14}$ & $8.8 \times 10^{13}$  \\
SCSO~J233951.1--551331.3 & 0.73 & $ 11.6\pm  3.9$ & $  9.9\pm  3.3$ & $ 8.8 \times 10^{11}\pm1.8 \times 10^{11}$ & $1.1 \times 10^{14}$ & $1.3 \times 10^{14}$  \\
SCSO~J233720.2--562115.1 & 0.75 & $ 10.7\pm  4.0$ & $  7.7\pm  2.9$ & $ 5.0 \times 10^{11}\pm1.2 \times 10^{11}$ & $8.5 \times 10^{13}$ & $7.0 \times 10^{13}$  \\
\enddata
\label{tab:data2}
\tablecomments{Catalog of the optical clusters with mass estimates
  $<3\times 10^{14} M_{\sun}$ from the $M(L_{200})$ values. Each cluster's 
  redshift is the mean photometric redshift computed using the elliptical 
  in the center of the cluster. The ID is based on the position of the
  BCG.}
\end{deluxetable*}

\section{Results and Conclusions}

In this article we have laid out the techniques and methods for our
analysis of a large multi-band optical survey with the Blanco
telescope and Mosaic-II instrument under the aegis of the Southern
Cosmology Survey.  We have obtained sub-arcsecond astrometric
precision and sufficient photometric accuracy for the estimation of
redshifts to $\delta z < 0.1$.  Forty-two optical cluster candidates
were identified from an area of the sky covering $\simeq$8~deg$^2$;
the richness and integrated galaxy luminosity of the clusters were
measured.  Based on correlations between these optical observables and
cluster mass as established by SDSS cluster surveys \citep{Johnston07,
  Reyes08} we provide mass estimates (in addition to positions and
redshifts) for 8 new clusters whose inferred masses lie above $3\times
10^{14} M_{\sun}$. These clusters are all likely to be detected by ACT
and SPT if these experiments reach their expected sensitivity levels.
Although the uncertainties on the estimated mass and inferred SZE
signal are large (factors of 2 or so), the accuracy of our cluster
positions and redshifts are quite good and typical for 4-band imaging
survey data.
Moreover, it is worth noting that the $Y-M$ relation varies quite a
bit from simulation to simulation (differences of a factor of 2 in $M$
at fixed $Y$), so our predicted masses are well within the uncertainty
range of the latest theoretical predictions. In an effort to reduce
the mass errors we have begun to estimate weak lensing masses from the
current Blanco data \citep{McInnes09}. We also note that significant
additional areas of the BCS are now publicly available and will be
presented in future publications.

The strength of the SCS is its multi-wavelength aspect and large sky
area coverage.  The 23hr region of the sky analyzed here has now been
surveyed in the UV by GALEX and in the X-ray band by {\it XMM-Newton};
over the next several months as these data become available, we will
incorporate them into the SCS.  Correlation analyses of these
multi-wavelength data should allow us to reduce the large errors on
inferred cluster masses and study the cluster selection biases across
wavebands.  Secure confirmation of all candidates, as well as
additional mass constraints, rest on follow-up optical spectroscopy,
which we are pursuing at the Southern African Large Telescope and
elsewhere.  Finally, once ACT and SPT data become available, the rich
optical data set analyzed here will be a valuable source for
understanding and quantifying the impact of large scale structure on
secondary anisotropies in the CMB.

\acknowledgments

FM would like to thank Txitxo Benitez for discussions and the use of BPZ
priors and templates. 
JPH would like to thank Kent Patterson for help during the early
phases of this project and the Aresty Center for Undergraduate
Research at Rutgers for financial support. AK has been supported by
NSF grant AST-0546035. LI thanks Centro de Astrofisica Fondap.
We thank Armin Rest and Frank Valdes for helpful discussion on the
image analysis and we also acknowledge David Spergel for several
important insights and suggestions.
We are thankful to Wayne Barkhouse, Mark Brodwin, Will High, Yen-Ting
Lin and the entire BCS team for the careful planning and execution of
the NOAO observations on which this paper is based.
Financial support was provided by the NASA LTSA program (grant number
NAG5-11714) and the National Science Foundation under the PIRE program
(award number OISE-0530095).

\end{document}